\documentclass [12pt] {report}
\usepackage[cp1251]{inputenc}
\usepackage[russian]{babel}
\usepackage[dvips]{graphicx}
\usepackage{graphicx,psfrag}
\usepackage{amssymb}
\usepackage{amsmath}
\newcommand {\eqdef} {\stackrel{\rm def}{=}}

\newcommand {\e} {\mathop{\rm e}\nolimits}
\newcommand {\sign} {\mathop{\rm sign}\nolimits}

\newcommand {\D}[2] {\displaystyle\frac{\partial{#1}}{\partial{#2}}}
\newcommand {\DD}[2] {\displaystyle\frac{\partial^2{#1}}{\partial{#2^2}}}

\newcommand {\al} {\alpha}

\newcommand {\La} {\Lambda}
\newcommand {\ka} {\varkappa}
\newcommand {\Ka} {\mbox{\Large$\varkappa$}}

\newcommand {\si} {\sigma}
\newcommand {\Si} {\Sigma}

\newcommand {\ga} {\gamma}

\newcommand {\de} {\delta}
\newcommand {\De} {\Delta}
\newcommand {\om} {\omega}
\newcommand {\iy} {\infty}
\newcommand {\prtl} {\partial}
\newcommand {\fr} {\displaystyle\frac}
\newcommand {\suml} {\sum\limits}

\newcommand {\wt} {\widetilde}

\newcommand {\be} {\begin{equation}}
\newcommand {\ee} {\end{equation}}
\newcommand {\ba} {\begin{array}}
\newcommand {\ea} {\end{array}}
\newcommand {\bp} {\begin{picture}}
\newcommand {\ep} {\end{picture}}
\newcommand {\bc} {\begin{center}}
\newcommand {\ec} {\end{center}}

\newcommand {\bt} {\begin{tabular}}
\newcommand {\et} {\end{tabular}}
\newcommand {\lf} {\left}
\newcommand {\rg} {\right}

\newcommand {\nin}{\noindent}

\newcommand {\cA} {{\cal A}}

\newcommand {\cD} {{\cal D}}
\newcommand {\cE} {{\cal E}}
\newcommand {\cF} {{\cal F}}

\newcommand {\cI} {{\cal I}}
\newcommand {\cJ} {{\cal J}}
\newcommand {\cK} {{\cal K}}
\newcommand {\cM} {{\cal M}}
\newcommand {\cO} {{\cal O}}
\newcommand {\cP} {{\cal P}}

\newcommand {\cR} {{\cal R}}
\newcommand {\cS} {{\cal S}}

\newcommand {\cW} {{\cal W}}
\newcommand {\bk} {{\bf k}}

\newcommand {\bq} {{\bf q}}
\newcommand {\bP} {{\bf P}}

\newcommand {\bR} {{\bf R}}

\newcommand{\mt}{\mbox{$|{\bf t}|$}}

\newcommand{\mP}{\mbox{$|{\bf P}|$}}

\newcommand{\mR}{\mbox{$|{\bf R}|$}}
\newcommand{\mr}{\mbox{$|{\bf r}|$}}
\newcommand{\Og}{\ensuremath{O_1(g)}}
\newcommand{\Ogg}{\ensuremath{O_{\ge2}(g)}}
\newcommand {\ses} {\medskip}
\newcommand{\bvec}[1]{{\bf #1}}
\newcommand {\pgbrk}{\pagebreak}

\def\2#1#2#3{{#1}_{#2}\hspace{0pt}^{#3}}
\def\3#1#2#3#4{{#1}_{#2}\hspace{0pt}^{#3}\hspace{0pt}_{#4}}
\newcounter{sctn}
\def\sec#1.#2\par{\setcounter{sctn}{#1}\setcounter{equation}{0}
                  \noindent{\bf\boldmath#1.#2}\bigskip\par}
\begin {document}

\ses\ses

\begin {titlepage}

\nin

\vspace{0.1in}

\begin{center}
{\large \bf Finsleroid--Relativistic Time--Asymmetric Space and  Quantized Fields
}\\
\end{center}

\vspace{0.3in}

\begin{center}

\vspace{.15in}
{\large G. S. Asanov\\}
\vspace{.25in}
{\it Division of Theoretical Physics, Moscow State University\\
119992 Moscow, Russia\\
(e-mail: asanov@newmail.ru)}
\vspace{.05in}

\end{center}


\ses

For well-defined
 Finsleroid-relativistic  space
$\cE_g^{SR}$ (with the upperscript SR meaning  Special-Relativistic)
 due only to accounting a  characteristic parameter $g$ which
  measures the deviation of the
  geometry from its
pseudoeuclidean precursor,
the creation of the respective quantization programs for relativistic physical
fields seems to be an urgent task.
The parameter  may take on the values over all the real range;
at $g=0$  the space is reduced to become an  ordinary pseudoeuclidean one.
In the present work, the formulation of theory for relativistic physical fields in
such a space is initiated.
A general method to solve respective scalar, electromagnetic, and spinor field equations
is proposed basing on the conformal flatness.
At any value of the parameter,
 the expansion of the relativistic fields  with respect to non-plane waves appeared is found,
which proposes a base upon which the fields can be quantized in context of  the
Finsleroid-relativistic  approach.
Remarkably, the regulators can naturally be  proposed to overcome divergences in
relativistic field integrals.
The respective key and basic concepts involved
 are presented.
For short, the abbreviations FMF, FMT, and FHF will be used for the
Finsleroid-relativistic  metric function,  the associated  metric
tensor, and the associated  Hamiltonian function, respectively.

\end{titlepage}

\ses\ses

\setcounter{sctn}{1} \setcounter{equation}{0}

{\nin\bf 1. Introduction}

\ses\ses

Choosing the basic metric function to be of the pseudoeuclidean form
is the geometric base proper for the nowadays relativistic field theory.
The Finsler-geometry methods
[1-2] may suggest the possibility  of extending the theory by using  metric functions of
 essentially more general types.
Below, we develop the  Finslerian $\cE^{SR}_g$--space approach
 specified by a characteristic parameter $g$.
Conceptually,  our interpretation of the possibility
 is based upon treating the centered
 $\cE^{SR}_g$--space
as an
Observation Space ($\cO\cS$ for short) for the quantum processes.
Some value of
$g$
is implied to be attached to
$\cO\cS$.\\

The notation $R$ will be used for the four-dimensional
vectors that issue from the  centre point $``O"$ and belong to the space: $R\in \cO\cS$.
The parameter $g$ is independent of the vector argument $R$ (is a constant over the whole
$\cO\cS$).
For the sake of convenience in process of calculations, the space
$\cO\cS$ or $\cE^{SR}_g$
 will be referred to a rectilinear coordinate system,
 which centre being identified with the  $``O"$,
and the components of the vectors with respect to the system  will be used:
$
R=\{R^p\}= \{R^0,R^1,R^2,R^3\}\equiv \{T,X,Y,Z\}.
$

We use a particular special-relativistic FMF
$F=F(g;T,X,Y,Z)$
subject to the following attractive conditions:

\ses\ses

\nin
 (P1) {\it The indicatrix-surface $\cI_g$ defined by the equation
$F(g;T,X,Y,Z)=1$
is a regular space of a
constant negative
curvature,}
 $R_{\text{Ind}}$,

\ses\ses

\nin
which is a convenient stipulating to seek for the nearest
 Riemannian-to-Finslerian relativistic  generalizations;
\ses

\nin
 (P2)
{\it The FMF is compatible with the principle of spatial isotropy} ({\it P}--{\it parity});

\ses\ses

\nin
 (P3) {\it The associated FMT is of the time-space signature $(+---)$};

\ses\ses



\nin
 (P4) {\it The principle of correspondence holds true},
\ses

\nin
 that is, the associated FMT
reduces exactly to its ordinary known special-- or general--relativistic prototype when
$R_{\text{Ind}}\to-1$, which physical  significance is quite transparent.
The limit value $R_{\text{Ind}}=-1$ corresponds  precisely to $g=0$, so that
$F(0;T,X,Y,Z)=\sqrt{T^2-X^2-Y^2-Z^2}$.

\ses
\ses

 All the items (P1)--(P4) are obeyed whenever one makes the choice
(A.12) (see Appendix A).
Vice versa, we can claim the following

\ses\ses

 { THE UNIQUENESS THEOREM.}
 \it The properties \rm(P1)--(P4), \it when treated as conditions imposed on
 the FMF, specify it unambiguously in the form
given by the definition
\rm(A.12).
 \rm

\ses
\ses
 \nin
The indicatrix curvature value is given by (A.29).

\ses\ses

NOTE. The $\cE_g^{SR}$--space is the Finsleroid-type relativistic
space founded upon retaining the
$P$--parity at the expense of violating the $T$--parity.
The resultant FMF $F$  proves to be  {\it $T$-asymmetric}.
 The associated FMT exhibits the 6-parametrical invariance.
 Various aspects of the space were formulated earlier [3-5].
 There exists an alternative space in which the $T$--parity holds under violating
 the $P$-- parity, assuming existence of a preferred geometrically-distinguished
 spatial direction [6].

\ses\ses

The basic properties of the $\cE^{SR}_g$--space approach to the theory of relativistic fields
can readily  be seen and exercised in case of the scalar field
(Section 2).
A novel qualitative feature is that the conservation laws cease being integrable,
for they  involve now the connection coefficients $\3Cpqr$
(see (2.17)--(2.18)).
In Section 3, the electromagnetic field is tackled with. In
 Section 4,  the spinor field equations are generalized
 accordingly.
In analyzing the field equations obtained, it is convenient to convert the
consideration into the quasi-pseudoeuclidean space.
Solutions to the  equations
prove to be direct images of solutions to the equations for
quasi-pseudoeuclidean fields.
 In the $O_1(g)$--approximation
(which is  accounting for the Finslerian corrections
 proportional to degrees of the  characteristic parameter $g$
of the order not higher than the first one)
the latter equations are of the ordinary pseudoeuclidean
 forms, so that  for them
the ordinary relativistic plane- wave expansion can be applied.
The fact is that in the $O_1(g)$--approximations
the form of the quasi-pseudoeuclidean metric tensor
is merely of the pseudoeuclidean type
(corrections are proportional to $g^2$).

However, the possibility of solving the $\cE_g^{SR}$--space relativistic field equations
in the general case not adhering at the $O_1(g)$--approximation
is not obvious.
In Section 5 the method is proposed to rigorously
solve the $\cE_g^{SR}$--space electromaganetic
field equations basing upon the remarkable fact that the respective Lagrangian
density is conformally invariant.
In Section 6 we show how the Lagrangians for scalar and spinor fields can be extended
to entail upon appropriate changes of variables the field equations of
 ordinary pseudoeuclidean form.
To this end, the concepts of conformal scalar and spinor fields are introduced.
Therefore, we obtain the total method to solve the
$\cE_g^{SR}$--equations for all the electromagnetic, scalar, and spinor fields.
Particularly,
the $\cE_g^{SR}$--space  versions of the Coulomb law and the Yakawa potential can
 explicitly be proposed.

Also, we find relevant extensions of plane-wave decompositions (as well as of
spherical-type solutions).
This circumstance opens up due vistas  to quantize
 $\cE_g^{SR}$--space fields in a convenient and rigorous way
 (Section 7).
The waves are not plane but conformally plane and the expansion of fields with respect to such
waves is offered.
The  operator of the four-dimensional momentum can readily be indicated for such waves.


The   $\cE_g^{SR}$--space  approach proposed in the present work
may, in principle, provide one with
simple explanation of the nature of ultraviolet as well as angular
divergences (which are
characteristic of ordinary Lorentz-invariant theory of quantized fields),
for the limiting transition with $g\to0$ in the    $\cE_g^{SR}$--space extended
integrals
to the integrals of the Lorentz-invariant theory proves to be far from being always possible!
There appear convenient regulators, the $\hat\cE_g^{SR}$--space weights,
to perform non-singular integrations over momenta
(Section 8).

Appendices A and B are devoted to explaining the  basic properties of the $\cE_g^{SR}$--space.
Appendix C describes and visualizes the behaviour of fronts of
respective non-plane  waves.
Appendices D and E present the conformal properties and various $O_1(g)$--approximations.

The pseudoeuclidean metric tensor will be presented by the components
$e_{ij}$, so that
$e_{ij}=e^{ij}={\rm diag}(1,-1,-1,-1)$;
the short derivative designations $\partial_i=\partial/\partial t^i $
and $\partial_p=\partial/\partial R^p $ will be used.


\ses\ses

\setcounter{sctn}{2} \setcounter{equation}{0}

{\nin\bf 2.   $\cE_g^{SR}$--space extension of Klein-Gordon
equation}

\ses \ses

Let
 $ \phi=\phi(R) $
  be a complex scalar field, so that
\be
\phi(R)= \phi_1(R)+i\phi_2(R), \qquad \phi^*(R)=
\phi_1(R)-i\phi_2(R),
\ee
 where
  $ \phi_1(R) $ and $ \phi_2(R) $
are two real scalar functions;
the star (*) means the complex conjugation.
 Introducing also the notation
$
\phi_p=\partial_p\phi $
and
$ \phi^*_p=\partial_p\phi^* $
 for
partial derivatives with respect to the  argument $R^p$, we
construct the Lagrangian
\be
 L_{\phi} =
g^{pq}(g;R)\phi^*_p(R)\phi_q(R)-m^2\phi^*(R)\phi(R)
\ee
  ($m$
denotes the rest mass of the scalar particle) and the Lagrangian
density
\be \La_{\phi}= L_{\phi}J,
\ee
where
\be J~:=
\sqrt{|\det(g_{pq}(g;R))|}=(j(g;R))^4
\ee
 denotes the Jacobian for
the FMT
 (see (A.18)),
  to use the action
integral
 \be S\{\phi\}=\int\La_{\phi}d^4R.
  \ee
   The associated
Euler-Lagrange derivatives
 \be
\cE_{\phi}~:=\partial_q\D{\La_{\phi}}{\phi_q}
-\D{\La_{\phi}}{\phi}, \qquad
\cE^*_{\phi}~:=\partial_q\D{\La_{\phi}}{\phi^*_q}
-\D{\La_{\phi}}{\phi^*}
 \ee
  can be found from Eqs. (2.1)--(2.6) to
be
 \be \fr1{J}\cE_{\phi}=\Box\phi^*+m^2\phi^*, \qquad
\fr1{J}\cE^*_{\phi}=\Box\phi+m^2\phi,
\ee
where
 \be
\Box=\fr1{J}\partial_p\Bigl(Jg^{pq}\partial_q\Bigr).
\ee
Since $\delta S=0\,\Longrightarrow\,\cE_{\phi}=0$, we get the
following
 {\it  $\cE_g^{SR}$--space scalar field equation}:
\be
\Box\phi+m^2\phi=0.
\ee



The respective
{\it $\cE_g^{SR}$--space scalar field current density}
\be
\cJ_p~:=  i(\phi^*\partial_p\phi-\phi\partial_p\phi^*)J
\ee
is conserved over solutions to the equation
 (2.9),
according to
\be
\partial_p\cJ^p=0,
\ee
where $\cJ^p=g^{pq}\cJ_q$.
Therefore the charge is still conserved
in the ordinary sense:
\be
Q=\int{\cal J}^0 d^3\bR
\ee
and
\be
\partial_0Q=-\int{\cal J}^a d\Si_a.
\ee

Considering
the respective energy-momentum tensor of scalar field:
\be
T_p{}^q~:=(\phi_p\D{\La_{\phi}}{\phi_q}
+\phi^*_p\D{\La_{\phi}}{\phi^*_q}
-\delta_p^q\La_{\phi})/J,
\ee
we obtain
\be
T_p{}^q=\phi_p\phi^{*q}+\phi^*_p\phi^q-\delta_p^qL_{\phi},
\ee
where
\be
\phi^q(R)=g^{qp}(g;R)\phi_p(R),
\qquad
\phi^{*q}(R)=g^{qp}(g;R)\phi^*_p(R).
\ee
Now, the $\cE_g^{SR}$--space extension of  the conservation law for
the
scalar field
energy-momentum tensor
reads
\be
\cD_qT_p{}^q~:=\partial_qT_p{}^q-C_p{}^r{}_qT_r{}^q+C_t{}^q{}_qT_p{}^t
\ee
and
\be
\cD_qT_p{}^q=0,
\ee
\ses
what can be verified by direct calculations;
$C_p{}^r{}_q$ are coefficients (A.20).



Let us apply
the quasi-pseudoeuclidean
transformation
$t^i=\si^i(g;R)$
(see (B.1)--(B.2))
and introduce the transformed scalar field in accordance with
$\phi(R)=u\lf(\si(g;R)\rg)$,
so that
\be
u(t)=\phi(R).
\ee
We call the field
$u(t)$
the
 {\it
 quasi-pseudoeuclidean
image of the scalar field}
$\phi(R)$.
We get
\be
\La_{\phi}=\La_u,
\ee
\ses
\be
L_u=
n^{ij}(g;t)u^*_i(t)u_j(t)-m^2u^*(t)u(t),
\ee
and
\be
\La_{u}=L_uh^{-3},
\ee
where the equality (B.19) has been used;
the tensor $n^{ij}(g;t)$
is given by (B.16).
The action integral
(2.5) goes over in
\be
S\{u\}=\int\La_{u}d^4t.
\ee
From
$\delta S=0\,\Longrightarrow\,\cE_{u}=0$
we obtain
the {\it
quasi-pseudoeuclidean
scalar field equation}:
\be
\Box u+m^2u=0,
\ee
where
\be
\Box=\partial_i(n^{ij}\partial_j).
\ee



Considering
the {\it
quasi-pseudoeuclidean
image of
the density of four-dimensional Finslerian scalar current}
\be
\cJ_i~:=  i(u^*\partial_iu-u\partial_iu^*)h,
\ee
we arrive at the conclusion
 that the current is conserved over solutions to the
equations (2.24):
\be
\partial_i\cJ^i=0,
\ee
where
$\cJ^i=n^{ij}\cJ_j$.

The energy-momentum tensor
\be
T_i{}^j~:=
\Bigl[
u_i\D{\La_u}{u_j}
+u^*_i\D{\La_u}{u^*_j}
-\delta_i^j\La_u\Bigr]/h
\ee
can be rewritten as
\be
T_i{}^j=u_i(t)u^{*j}(t)+u^*_i(t)u^j(t)-\delta_i^jL_u,
\ee
where
\be
u^j(t)=n^{ji}(g;t)u_i(t),
\qquad
u^{*j}(t)=n^{ji}(g;t)u^*_i(t).
\ee
From
(2.24) the following conservation law follows for the
quasi-pseudoeuclidean
tensor (2.29):
\be
\cD_jT_i^j~:=\partial_jT_i^j-N_i{}^m{}_jT_m^j+N_n{}^j{}_jT_i^n
\ee
and
\be
\cD_jT_i^j=0,
\ee
where
$N_i{}^m{}_j$ ---
the connection coefficients of
the quasi-pseudoeuclidean
space
(see (B.26)).


Thus we have proved

\ses

{ PROPOSITION 1.} {\it Solutions to the Finslerian scalar field equations
{\rm (2.9)} are
the quasi-pseudoeuclidean
images of solutions to the
equations
{\rm (2.24)}
for  quasi-pseudoeuclidean
scalar field.
}

\ses

The transformations
\be
\cJ^p(R)=R^p_i(g;t)\cJ^i(t)
\ee
\ses
and
\be
\cJ^i(t)=t^i_p(g;R)\cJ^p(R)
\ee
together with
\be
T^{pq}(R)=R^p_i(g;t)R^q_j(g;t)T^{ij}(t)
\ee
\ses
and
\be
T^{ij}(t)=t^i_p(g;R)t^j_q(g;R)T^{pq}(R)
\ee
are implied;
 $t^i_p$
and
 $R^p_i$
 are the coefficients
 (B.9) and (B.13).



{\it
In the
$O_1(g)$--approximation}
we have simply
$
n_{ij}=e_{ij}$ and $ n^{ij}=e^{ij}.$
The coefficients $N_i{}^k{}_j$
are proportional to $g^2$ (see (B.26))
and, therefore,  disappear at the
$O_1(g)$--level.
With these restrictions, the equation
 (2.24)--(2.25)
reduces to the ordinary Klein-Gordon form, so that
 we are entitled to apply the ordinary
plane-wave expansion to the
  quasi-pseudoeuclidean
image of scalar field:
\be
u(t)= \fr1{(2\pi)^{3/2}}\int e^{ik_jt^j}u(k)\delta(k^2-m^2)d^4k
\ee
and
\be
u(t)=u^+(t)+u^-(t)
\ee
with
\be
u^+(t)
= \fr1{(2\pi)^{3/2}}\int e^{ik_jt^j}u^+(k)
\delta(k^2-m^2)d^4k,
\qquad k_0>0,
\ee
\bigskip
\be
u^-(t)
= \fr1{(2\pi)^{3/2}}\int e^{-ik_jt^j}u^-(k)
\delta(k^2-m^2)d^4k,
\qquad k_0>0,
\ee
\bigskip
\be
(u^+(k))^*=u^{*-}(k),
\qquad
(u^-(k))^*=u^{*+}(k),
\ee
\bigskip
\be
u^+({\bf k})=\fr{u^+(k)}{\sqrt{2k_0}}
,\qquad
u^-({\bf k})=\fr{u^-(k)}{\sqrt{2k_0}},
\ee
\bigskip
\be
u^+(t)
= \fr1{(2\pi)^{3/2}}\int e^{ik_jt^j}u^+({\bf k})
\fr{d{\bf k}}{\sqrt{2k_0}},
\ee
\bigskip
\be
u^-(t)
= \fr1{(2\pi)^{3/2}}\int e^{-ik_jt^j}u^-({\bf k})
\fr{d{\bf k}}{\sqrt{2k_0}},
\ee
\ses\\
where
$k_0=\sqrt{{\bf k}^2+m^2}>0$.
 \ses
The energy-momentum conservation law  (2.31)--(2.32)
takes on the ordinary integrable form
\be
\partial_jT_i^j=0.
\ee

From the above we obtain
\ses

{ PROPOSITION 2}. \it At
the
$O_1(g)$--approximation level
of consideration
the  quasi-pseudoeuclidean
image
 \rm (2.19) \it
of scalar field
is reduced to
merely
pseudoeuclidean
case.
\rm
\ses




Returning back from the
  quasi-pseudoeuclidean
space to the initial $\cE^{SR}_g$--space,
 we observe that the plane-wave decomposition should be
replaced by the non-plane-wave decomposition:
\be
\phi(R)=\phi^+(R)+\phi^-(R), \qquad
\phi^*(R)=\phi^{*+}(R)+\phi^{*-}(R)
\ee
\ses
with the functions
\be
\phi^+(R)
= \fr1{(2\pi)^{3/2}}\int e^{ik_n\si^n(g;R)}\phi^+(k)
\delta\lf(k^2-m^2\rg)d^4k
\ee
\bigskip
and
\be
\phi^-(R)
= \fr1{(2\pi)^{3/2}}\int e^{-ik_n\si^n(g;R)}\phi^-(k)
\delta\lf(k^2-m^2\rg)d^4k.
\ee
The ordinary conjugation properties remain valid:
\be
(\phi^+(k))^*=\phi^{*-}(k),
\qquad
(\phi^-(k))^*=\phi^{*+}(k).
\ee

Thus,
the $\cE^{SR}_g$--space
waves of scalar particles are not plane unless $g=0$.
On introducing the functions
\be
u^+(\bk)=\phi^+({\bf k})=\fr{\phi^+(k)}{\sqrt{2k_0}}
,\qquad
u^-(\bk)=\phi^-({\bf k})=\fr{\phi^-(k)}{\sqrt{2k_0}},
\ee
we get the representations
\be
\phi^+(R)
= \fr1{(2\pi)^{3/2}}\int e^{ik_n\si^n(g;R)}\phi^+({\bf k})
\fr{d^3{\bf k}}{\sqrt{2k_0}}
\ee
and
\be
\phi^-(R)
= \fr1{(2\pi)^{3/2}}\int e^{-ik_n\si^n(g;R)}\phi^-({\bf k})
\fr{d^3{\bf k}}{\sqrt{2k_0}}.
\ee





\ses\ses


\setcounter{sctn}{3}
\setcounter{equation}{0}
{\nin\bf 3. $\cE_g^{SR}$--space extension of  electromagnetic field equations}

\ses\ses

Let a covariant field
$
\cA=\{A_p(R)\}
$
be an electromagnetic vector potential.
We shall use the partial derivatives
$
A_{p,q}=\partial_qA_p
$
and, as usually, construct
the {\it
tension tensor of electromagnetic field}
\be
F_{pq}=A_{p,q}-A_{q,p}
\ee
and introduce the concomitant tensors
\be
F_p{}^q=g^{qr}F_{pr},
\qquad
F^{pq}=g^{pr}g^{qs}F_{rs}.
\ee
After that, we can introduce
the {\it
density of the $\cE^{SR}_g$--space induction tensor}
\be
\cD^{pq}(R)~:= J(g;R)g^{pt}(g;R)g^{qs}(g;R)F_{ts}(R)
\ee
and use the direct extension
\be
\La_{A}=
 -\fr14\cD^{pq}F_{pq}
\ee
of the classical (rescaled)
Lagrangian density
of
electromagnetic field
to set forth the action integral
\be
S\{A\}=\int\La_{A}d^4R.
\ee
The associated Euler-Lagrange derivatives
\be
\cE^p_{\cA}~:= \partial_q\D{\La_{A}}{A_{p,q}}=
2\partial_q\D{\La_{A}}{F_{qp}}
\ee
can be found from (3.1)--(3.5) to read
\be
\cE^p_{A}=\partial_q \cD^{pq}.
\ee
Since
 $ \delta S=0\,\Longrightarrow\,\cE^p_{A}=0$,
we obtain
the
{\it
$\cE^{SR}_g$--space electromagnetic field equations}
\be
\partial_q \cD^{pq}=0
\ee
(in vacuum).
Additionally the cyclic equations
\be
\partial_r{F_{pq}}
+
\partial_q{F_{rp}}
+
\partial_p{F_{qr}}
=0
\ee
are fulfilled.
The corresponding
{\it
energy-momentum tensor of the $\cE^{SR}_g$--space electromagnetic field}
\be
T_p{}^q~:= -F_{pr}F^{qr}+\fr14\delta_p{}^qF_{st}F^{st}
\ee
will satisfy  the conservation law of the form
 (2.17)--(2.18).



Let us introduce
the   {\it
quasi-pseudoeuclidean image for the initial
electromagnetic vector potential
$\cA$}:
\be
U_i(t)~:= R^p_i(g;t)A_p(\mu(g;t))
\ee
and, after that, introduce the corresponding
definitions for the tension tensor
\be
f_{ij}(t)~:= R^p_i(g;t)R^q_j(g;t)F_{pq}(R)
\ee
together with the contravariant tensor
\be
f^{ij}(t)~:= t^i_p(g;R)t^j_q(g;R)F^{pq}(R).
\ee
\ses
The inverse relations take the form
\be
A_p(R)=t_p^i(g;R)U_i(t),
\ee
\bigskip
\be
F_{pq}(R)=t_p^i(g;t)t_q^j(g;t)f_{ij}(t),
\ee
\bigskip
\be
F^{pq}(R)=R^p_i(g;t)R^q_j(g;t)f^{ij}(t),
\ee
\bigskip
\be
f^{ij}(t)=n^{im}n^{jn}f_{mn}(t),
\ee
\ses
and the representation
\be
f_{mn}(t)=\D{U_n(t)}{t^m}-\D{U_m(t)}{t^n}
\ee
is valid.


Next, we introduce the Lagrangian density
\be
\La_{A}=
\La_U,
\ee
where
\be
\La_{U}=L_Uh^3
\ee
and
\be
L_U=-\fr14n^{ij}n^{mn}f_{im}(t)f_{jn}(t)\equiv-\fr14f^{im}(t)f_{im}(t),
\ee
so that the action integral
(3.5)
transforms to
\be
S\{U\}=\int\La_Ud^4t.
\ee
From the action principle
$\delta S=0$
the equations of the ordinary form
\be
\D{f^{in}}{t^i}=0
\ee
ensue.
Besides,
the counterpart of (3.9) is valid:
\be
\D{f_{mn}}{t^i}+
\D{f_{im}}{t^n}+
\D{f_{ni}}{t^m}=0
\ee
The conservation law of the form
(2.31)--(2.32)
is applicable to the energy-momentum tensor
\be
T_i^j=
-f^{jm}f_{im}+
\frac14\delta^i_jf^{mn}f_{mn}.
\ee
Inserting any solution
{$U_i=U_i(t)$}
to the equation (3.23)--(3.24)
in the right-hand part of the relation
 (3.14) yields a field $A_p(R)$,
which gives a solution to the initial  $\cE_g^{SR}$--space
electromagnetic field equations
(3.8).


{\it
In
the $O_1(g)$--approximation}
the equations (3.23) reduce to the ordinary
Maxwell-type equations:
\be
\partial_jf^{ij}=0
\quad\mbox{with the functions}\quad
f^{ij}=e^{im}e^{jn}f_{mn},
\ee
and we can use the known solutions formed by plane waves:
\be
U_i(t)=\frac1{(2\pi)^{3/2}}\int e^{ik_jt^j}U_i(k)\delta(k^2)d^4k,
\ee
or in the form
\be
U_i(t)=U_i^+(t)+U_i^-(t)
\ee
\ses
with the functions
\be
U_i^{\pm}(t)=\frac1{(2\pi)^{3/2}}\int e^{\pm
ik_jt^j}U_i^{\pm}(\bvec k)\frac{d\bvec k}{\sqrt{2k_0}},
\ee
where $k_0=\sqrt{{\bk}^2}$.
Returning into the initial space, we get
\be
A_p(R)=A_p^+(R)+A_p^-(R)
\ee
\ses
with the functions
\be
A_p^{\pm}(R)=t^i_p(g;R)U_i^{\pm}(t)=\frac1{(2\pi)^{3/2}}t^i_p(g;R)\int
e^{\pm ik_n\si^n(g;R)}U_i^{\pm}(\bk)\frac{d\bk}{\sqrt{2k_0}}\,.
\ee
Ordinary plane-wave decomposition (3.27)
turn over into non-plane-wave  decomposition
\be
A_p(R)= \fr1{(2\pi)^{3/2}}\int e^{ik_n\si^n(g;R)}A_p(R,k)
\delta(k^2)d^4k,
\ee
where
\be
A_p(R,k)=t^i_p(g;R)U_i(k).
\ee

Whence we have arrived at the conclusion:

\ses\ses

{ PROPOSITION 3.} {\it
In
the $O_1(g)$--approximation,
solutions to the initial  $\cE_g^{SR}$--space electcromagnetic field
equations
{\rm (3.8)} present  straightforward images of solutions to
the linear
equations
{\rm (3.26)}
 of the ordinary Maxwell form}.

\ses\ses





\ses\ses

\setcounter{sctn}{4}
\setcounter{equation}{0}
{\nin\bf 4.   $\cE_g^{SR}$--space extension of spinor field equations}

\ses\ses

For the four-component spinor
$\psi(R)$
of the Dirac type
we introduce, in agreement with known methods of general-relativistic approaches,
the {\it
 $\cE_g^{SR}$--space
spinor derivatives}
\ses
\be
D_p\psi(R)=\prtl_p\psi(R)-Z_p(g;R)\psi(R),
\qquad
D_p\bar\psi(R)=\prtl_p\bar\psi(R)+\bar\psi(R)Z_p(g;R)
\ee
and,  respectively,
the
{\it
  $\cE_g^{SR}$--space
spinor connection coefficients}
\be
Z_p(g;R)=-\fr18{R^{PQ}}_p(g;R)(\ga_P\ga_Q-\ga_Q\ga_P),
\ee
where ${R^{PQ}}_p(g;R)$  --- the associated Ricci rotation coefficients
(see (A.37)--(A.39))
and the identity
\be
D_q\gamma^p=\cD_q\gamma^p-Z_q\gamma^p+\gamma^pZ_q=0
\ee
is fulfilled.
The notation
${\gamma}_P$ is used for the ordinary Dirac's matrices,
so that
\be
\ga_P\ga_Q+\ga_Q\ga_P=2a_{PQ},
\ee
where
$\{a_{PQ}\}={\rm diag}(1,-1,-1,-1)$;
we have also introduced the notation
\be
\gamma^p(g;R)=g^{pq}(g;R)e^P_q(g;R)\gamma_P,
\ee
where
$
e^P_q(g;R)
$
is the orthonormal invariant frame (see (A.35)).
The conjugation operation reads
\be
\bar\psi(R)
=
(\psi(R))^+\gamma^0(g;R).
\ee
Therefore, the  $\cE_g^{SR}$--extended Dirac-type equation reads
\be
-i\gamma^p(g;R)D_p\psi(R)+m\psi(R)=0
\ee
and the conjugated version reads
\be
i(D_p\bar\psi(R))\gamma^p(g;R)+m\bar\psi(R)=0;
\ee
here,
 $m$
stands for the rest mass of the spinor particle.
For the corresponding current density
\be
{\cal J}^p=J\bar\psi\gamma^p\psi
\ee
the direct calculations yield
$$
\partial_p{\cal
J}^p=
J\lf(\partial_p(\bar\psi)\gamma^p\psi+
\bar\psi\gamma^p\partial_p\psi+\bar\psi(\cD_p\gamma^p)\psi\rg)=
J\lf((D_p\bar\psi)\gamma^p\psi+\bar\psi\gamma^p(D_p\psi)\rg),
$$
where the property
(4.3)
has been used.
The spinor energy-momentum tensor is given by the formula
\be
T^{pq}=\frac i2\lf(\bar\psi\gamma^p
D^q\psi-(D^q\bar\psi)\gamma^p\psi\rg).
\ee
By virtue of the
equations (4.1)--(4.8)
we can readily infer the conservation laws of the form
 (2.11)--(2.13)
and
(2.17)--(2.18).


For the spinor
\be
v(t) = \psi(\mu(g;t)),
\ee
which arises on performing a
due transformation into
quasi-pseudoeuclidean space,
we get the equations which are similar to those shown above, namely,
\be
D_iv(t)=\prtl_iv(t)-T_i(g;t)v(t),
\qquad
D_i\bar v(t)=\prtl_i\bar v(t)+\bar v(t)T_i(g;t)
\ee
and we also get
respectively
the  {\it
$\cE_g^{SR}$--space
spinor connection coefficients}
\be
T_i(g;t)=-\fr18{R^{PQ}}_i(g;t)(\ga_P\ga_Q-\ga_Q\ga_P)
\ee
(${R^{PQ}}_i(g;t)$ stand for the quasi-pseudoeuclidean Ricci rotation coefficients;
see (B.34)),
from the definition of which the following counterpart of the identity (4.3)
ensues:
\be
D_j\gamma^i=\cD_j\gamma^i-T_j\gamma^i+\gamma^iT_j=0,
\ee
where
\be
\gamma^i(g;t)=n^{ij}(g;t)f^P_j(g;t)\gamma_P
\ee
with
$
f^P_j(g;t)
$
being the frame
(B.28)).
Also,
\be
\bar v(t)
=
(v(t))^+\gamma^0(g;t).
\ee
The Dirac-type equation in the quasi-pseudoeuclidean space takes on the form
\be
-i\gamma^i(g;t)D_iv(t)+mv(t)=0,
\ee
and for its conjugation we should write
\be
i(D_i\bar v(t))\gamma^i(g;t)+m\bar v(t)=0.
\ee
For the respective spinor current density
\be
{\cal J}^i=J\bar v\gamma^iv
\ee
and for the respective spinor energy-momentum tensor
\be
T^{ij}=\frac i2(\bar\psi\gamma^i
D^j\psi-(D^j\bar\psi)\gamma^i\psi)
\ee
the conservation laws of the form
(2.27)
and
(2.31)--(2.32)
 hold.


The fact that
in the quasi-pseudoeuclidean space the  $\cE_g^{SR}$--space
corrections are of the order of
$g^2$ opens up the direct possibility to solve
the considered spinor equations
{\it at the $O_1(g)$--approximation}
by using results of ordinary relativistic theory.
In particular, we can take
solutions to be of the plane-wave type:
\be
v(t)=\frac1{(2\pi)^{3/2}}\int
e^{ik_nt^n}\delta(k^2-m^2)v(k)d^4k,
\ee
or
\be
v(t)=v^+(t)+v^-(t)
\ee
with the functions
\ses
\be
v^{\pm}(t)=\frac1{(2\pi)^{3/2}}\int e^{\pm
ik_jt^j}v^{\pm}(\bvec k)d\bvec k,
\ee
where
\be
v^{\pm}(\bvec k)=\frac{\theta(k_0)v^{\pm}(k)}{2k^0}.
\ee
On returning back in the initial  $\cE_g^{SR}$--space,
 we obtain the representation
\be
\psi(R)=\frac1{(2\pi)^{3/2}}\int
e^{ik_n\si^n(g;R)}\delta(k^2-m^2)\psi(k)d^4k.
\ee


\ses\ses


\setcounter{sctn}{5}
\setcounter{equation}{0}
{\nin\bf 5. Conformal  method of solution for  $\cE^{SR}_g$--space
electromagnetic field equations}

\ses\ses




Given a vector field
$
U_i(t)
$
in the quasi-pseudoeuclidean space,
we perform the transformation
to new components
$
B_j(r)
$
according to the vector law
\be
 U_i(t)~:= k^j_iB_j(r)
\ee
which owing to (B.35)--(B.38)
reads explicitly as
\be
U_i=\fr1h\xi(B_i+\ga B_mn_in^m).
\ee
Next, we introduce the tensors
\be
B_{ij} ~:= \frac {\prtl B_j}{\partial r^i}
-
\frac{\prtl B_i}{\partial r^j},
\ee
\ses
\be
B_i{}^j~:= B_{im}e^{jm},
\qquad
B^{ij}~:= B_n{}^je^{ni},
\ee
and apply the transformation law
\be
f_{ij}= B_{mn}k^m_ik^n_j,
\ee
where
$f_{ij}$ are constructed according to the rule (3.18).
The concomitant tensors are
$
f_i{}^j~:= f_{im}n^{jm}
$
and
$
f^{ij}~:= f_n{}^jn^{ni}.
$

Due insertions yield explicitly
\be
f_{ij}=\fr1{h^2}\xi^2\Bigl[ B_{ij}+\ga h^2(B_{im}n_j-B_{jm}n_i)n^m\Bigr],
\ee
\ses
\be
f_i{}^j=\xi^2\Bigl[B_i{}^j-\ga h^2B^j{}_mn_in^m+\ga(2h^2+\fr1{h^2})\Bigr]B_{jm}n^jn^m,
\ee
and
\be
f^{ij}=
\xi^2\Bigl[h^2B^{ij}-\ga(\ga^3+ h^2+h)n^iB^j{}_mn^m
+\ga(2h^4+1)B^i{}_mn^jn^m\Bigr].
\ee



The equations
\be
\frac {\prtl B_{im}}{\partial r^j}
+
\frac{\prtl B_{ji}}{\partial r^m}
+
\frac{\prtl B_{mj}}{\partial r^i}
=0
\ee
are obviously valid. They  entail the electromagnetic field equations (3.24).
Also, straightforward calculations show that
the equations
\be
\frac {\prtl B^{ij}}{\partial r^j}=0
\ee
would entail the electromagnetic field equations (3.23) for the tensor (5.5).

Thus we have arrived at the following

\ses\ses

{PROPOSITION 4}.
The solutions to the electromagnetic field equations in the space $\cE^{SR}_g$
are the transforms of the solutions to
the ordinary Maxwell equations in accordance with the
formulae (3.14)--(3.15) and (5.6)--(5.10).

\ses
\ses

This result offers a handy way to export the electromagnetic field solutions
 from the
ordinary pseudoeuclidean space to the  $\cE^{SR}_g$--space
under study.



In particular, the ordinary plane-wave solution is now extending as follows:
\be
B_m=b_m\e^{i\Phi}, \qquad b_m = {\rm constants},
\ee
and
\be
U_j=a_j\e^{i\Phi},
\qquad
a_j = {\rm constants},
\ee
with
$\Phi$ being the phase (B.51).

Taking for definiteness
\be
k_n=(k_0,k_1=-k_0,0,0)
\ee
and
\be
b_i=(0,0,0,b_3), \quad b^i=(0,0,0,b^3),\quad b^3=-b_3,
\ee
we get for the vector potential
$
A_{\{k\}p}
$
the following explicit components:
\be
A_{\{k\}0}=
\lf(\fr g2(2h-1)-(h-1)k\rg)\fr{R^3}
{\mR}
\fr{j\ka b_3}
{L}
\e^{i\Phi},
\ee
\ses
\be
A_{\{k\}1}=
\lf(h-1-\fr12gk\rg)\fr{R^1R^3}
{\mR^2}
\fr{j\ka b_3}
{L}
\e^{i\Phi},
\ee
\ses
\be
A_{\{k\}2}=
\lf(h-1-\fr12gk\rg)\fr{R^2R^3}
{\mR^2}
\fr{j\ka b_3}
{L}
\e^{i\Phi},
\ee
\ses
\be
A_{\{k\}3}=
\Biggl[L+
\lf(h-1-\fr12gk\rg)
\fr{(R^3)^2}
{\mR^2}
\Biggr]
\fr{j\ka b_3}
{L}
\e^{i\Phi},
\ee
where
\be
L=1+gk-k^2
\ee
and (3.14) has been applied.
Contracting yields
the simple result
\be
R^pA_{\{k\}p}=hj\ka R^3b_3\e^{i\Phi}.
\ee
The notation
\be
k=\fr{R^0}
{\mR}
\ee
and
the conformal multipliers
$\ka$ and $\xi$
(see (B.36) and (B.43)) have conveniently been used.


When
\be
 R^0=0,
\ee
we have
\be k=0,
\quad L=1, \ee
 \ses
\be
F\bigl|_{R^0=0}
=(-g_-)^{G_+/2}(g_+)^{-G_-/2}\mR, \qquad
j\bigl|_{R^0=0}
=\lf(\fr{-g_-}{g_+}\rg)^{-G/4},
\ee
and the quantity $\Phi$ reduces to
\be
\Phi_0
=k_0C_2{\mbox{$|{\bf R}|$}}^{\frac12(h-1)}(-\frac G2{\mbox{$|{\bf R}|$}}-R^1).
\ee
We obtain
\be
A_{\{k\}0}\bigl|_{R^0=0}
=
\fr g2(2h-1)\fr{R^3}
{\mR}
{\mR^{\frac12(h-1)}}Cb_3
\e^{i\Phi_0},
\ee
\ses
\be
A_{\{k\}1}\bigl|_{R^0=0}
=
(h-1)\fr{R^1R^3}
{\mR^2}
{\mR^{\frac12(h-1)}}Cb_3
\e^{i\Phi_0},
\ee
\ses
\be
A_{\{k\}2}\bigl|_{R^0=0}
=
(h-1)\fr{R^2R^3}
{\mR^2}
{\mR^{\frac12(h-1)}}Cb_3
\e^{i\Phi_0},
\ee
\ses
\be
A_{\{k\}3}\bigl|_{R^0=0}
=
\Biggl[1+
(h-1)
\fr{(R^3)^2}
{\mR^2}
\Biggr]
{\mR^{\frac12(h-1)}}Cb_3
\e^{i\Phi_0}.
\ee
 $C$ and $C_2$ are constants.



For the contravariant components
$
A_{\{k\}}^p(R)=g^{pq}(g;R)A_{\{k\}q}(R)
$
calculations yield the results
\be
A_{\{k\}}^0=
(kh-f)\fr{R^3}
{\mR}
\fr{b^3}
{j\ka L}
\e^{i\Phi},
\ee
\ses
\be
A_{\{k\}}^1=
\lf(h-(k-g)f-L\rg)\fr{R^1R^3}
{\mR^2}
\fr{b^3}
{j\ka L}
\e^{i\Phi},
\ee
\ses
\be
A_{\{k\}}^2=
\lf(h-(k-g)f-L\rg)\fr{R^2R^3}
{\mR^2}
\fr{b^3}
{j\ka L}
\e^{i\Phi},
\ee
\ses
\be
A_{\{k\}}^3=
\Biggl[L+
\lf(h-(k-g)f-L\rg)
\fr{(R^3)^2}
{\mR^2}
\Biggr]
\fr{b^3}
{j\ka L}
\e^{i\Phi},
\ee
where
\be
f=k-\fr g2.
\ee


When
\be
 R^0=0,
\ee
we obtain
\be
A_{\{k\}}^0\bigl|_{R^0=0}
=
\fr g2\fr{R^3}
{\mR}
\fr1{\mR^{\frac12(h-1)}}Cb^3
\e^{i\Phi_0},
\ee
\ses
\be
A_{\{k\}}^1\bigl|_{R^0=0}
=
\lf(h-1-\fr{g^2}2\rg)\fr{R^1R^3}
{\mR^2}
\fr1{\mR^{\frac12(h-1)}}Cb^3
\e^{i\Phi_0},
\ee
\ses
\be
A_{\{k\}}^2\bigl|_{R^0=0}
=
\lf(h-1-\fr{g^2}2\rg)\fr{R^2R^3}
{\mR^2}
\fr1{\mR^{\frac12(h-1)}}Cb^3
\e^{i\Phi_0},
\ee
\ses
\be
A_{\{k\}}^3\bigl|_{R^0=0}
=
\Biggl[1+
\lf(h-1-\fr{g^2}2\rg)
\fr{(R^3)^2}
{\mR^2}
\Biggr]
\fr1{\mR^{\frac12(h-1)}}Cb^3
\e^{i\Phi_0}.
\ee


The associated electromagnetic field tension tensor can also be found explicitly, namely
\be
f_{ij}=i(k_ib_j-k_jb_i)e^{i\Phi}
\ee
(owing to (5.11) and (3.18))
and
\be
F_{pq}=i(\rho_p^nk_nA_{\{k\}q}-\rho_q^nk_nA_{\{k\}p})=
[(\rho_p^0-\rho_p^1)A_{\{k\}q}-(\rho_q^0-\rho_q^1)A_{\{k\}p}]k_0i
\e^{i\Phi}\ee
(the transformation (3.15) has been used)
with the components
\be
F_{01}=\Bigl[\fr g2-\ga(k-g)+(\ga h-\fr g2f)n^1\Bigr]
n^3
\fr{(j\ka )^2}{L}
b_3k_0i
\e^{i\Phi},
\ee
\ses
\be
F_{02}=(\ga h-\fr g2f)n^2n^3
\fr{(j\ka )^2}{L}b_3k_0i
\e^{i\Phi},
\ee
\ses
\be
F_{03}=(\ga h-\fr g2f)n^3n^3
\fr{(j\ka )^2}{L}b_3k_0i
\e^{i\Phi},
\ee
\ses
\be
F_{12}=
(\ga-\fr g2k)(n^1-n^2)n^3
\fr{(j\ka )^2}{L}b_3k_0i
\e^{i\Phi},
\ee
\ses
\be
F_{13}=
\Bigl[fk-h+(f-kh)n^1-(\ga-\fr g2k)n^1n^1\Bigr]n^3n^3
\fr{(j\ka )^2}{L}b_3k_0i
\e^{i\Phi},
\ee
\ses
\be
F_{23}=
\Bigl[f-kh-(\ga-\fr g2k)n^1\Bigr]n^2
\fr{(j\ka )^2}{L}b_3k_0i
\e^{i\Phi}.
\ee


When
\be
 R^0=0,
\ee
the above representations are reduced to read merely
\be
F_{01}\bigl|_{R^0=0}
=
\Bigl[\fr g2+\ga g+(\ga h+\fr{ g^2}4)n^1\Bigr]n^3
{\mR^{\frac12(h-1)}}Cb_3
\e^{i\Phi_0},
\ee
\ses
\be
F_{02}\bigl|_{R^0=0}
=
(\ga h+\fr{ g^2}4)n^2n^3
{\mR^{\frac12(h-1)}}Cb_3
\e^{i\Phi_0},
\ee
\ses
\be
F_{03}\bigl|_{R^0=0}
=
(\ga h+\fr{ g^2}4)n^3n^3
{\mR^{\frac12(h-1)}}Cb_3
\e^{i\Phi_0},
\ee
\ses
\be
F_{12}\bigl|_{R^0=0}
=
\ga (n^1-n^2)n^3
{\mR^{\frac12(h-1)}}Cb_3
\e^{i\Phi_0},
\ee
\ses
\be
F_{13}\bigl|_{R^0=0}
=
(- h-\fr g2n^1-\ga n^1n^1)n^3n^3
{\mR^{\frac12(h-1)}}Cb_3
\e^{i\Phi_0},
\ee
\ses
\be
F_{23}\bigl|_{R^0=0}
=
(-\fr g2-\ga n^1)n^2
{\mR^{\frac12(h-1)}}Cb_3
\e^{i\Phi_0}.
\ee


\ses\ses

NOTE.  Upon substituting (B.42) the Lagrangian density (3.4) takes on the form
\be
\cD^{pq}(R)=\sqrt{|\det(s_{pq}(g;R))|}\,s^{pt}(g;R)s^{qs}(g;R)F_{ts}(R)
\ee
that  corresponds to the flat-space case, for the curvature tensor constructed from
the conformal metric tensor $\{s_{pq}\}$ vanishes identically.
This is the reason why the above Proposition 4 is valid.
The respective transformation for the electromagnetic vector potential is
\be
A_p(R)=\rho_p^m(g;r)B_m(r)
\ee
with
\be
\rho_p^m=t^i_pk^m_i
\ee
(cf. (3.33) and (5.1)).

\ses\ses

At the $O_1(g)$--level of consideration,
the electromagnetic components (5.15)--(5.18) reduce to read
\be
A_{\{k\}0}=
\fr12 g\fr{\mR R^3}
{S^2}
{ b_3}
\e^{ik_pR^p},
\ee
\ses
\be
A_{\{k\}1}=
-\fr12g\fr{R^0}{\mR}\fr{R^1R^3}
{S^2}
{ b_3}
\e^{ik_pR^p},
\qquad
A_{\{k\}2}=
-\fr12g\fr{R^0}{\mR}\fr{R^2R^3}
{S^2}
{ b_3}
\e^{ik_pR^p},
\ee
\ses
\be
A_{\{k\}3}=\Biggl\{1-\fr12g\Biggr[i\mR k_0
+\fr12(1+ ik_aR^a)\ln\fr{R^0-\mR}{R^0+\mR}
+\fr{R^0}{\mR}
\fr{(R^3)^2}
{S^2}
\Biggr]
\Biggr\}
{ b_3}
\e^{ik_pR^p},
\ee
where
$
S^2=(R^0)^2-\mR^2$
 and (B.56) has been applied.

The {\it spherical waves of photons} are obtained upon taking
 the electromagnetic vector potential in the conformally flat space in the ordinary way:
\be
B_{\{\om jm\}0}=0
\ee
and
\be
B_{\{\om jm\}a}(r^q)=e^{ik_0r^0}
\int
B_{\{\om jm\}a}(\bk)e^{ik_br^b}\fr{d^3k}{(2\pi)^3}
\ee
with
\be
B_{\{\om jm\}a}(\bk)
=\fr{4\pi^2}{\om^{3/2}}\de(|\bk|-\om)
Y_{\{jm\}a}(\bk/\om),
\ee
where $\om=k_0$
and
$
Y_{\{jm\}a}
$
are required spherical functions subjected to the condition
$\lf(\bk {\bf Y}_{\{jm\}}(\bk/\om)\rg)=0$.
The quasi-pseudoeuclidean version reads
\be
U_{\{\om jm\}i}(t)
=
k^a_i(g;t)
B_{\{\om jm\}a}(r^q(g;t))
\ee
and returning back into the
  $\cE_g^{SR}$--space
  yields the components
\be
A_{\{\om jm\}p}(R)
=
\rho^a_p(g;R)
B_{\{\om jm\}a}(r^q(g;R)).
\ee
Since in the pseudoeuclidean approach proper the spherical electromagnetic waves
are expanding with the light velocity value, which relates to $(r^0)^2-{\bf r}^2=0$,
then in the initial
 $\cE_g^{SR}$--space
 we must respectively have $F(g;R)=0$, whence we may be certain that the respective velocity
 value in the  $\cE_g^{SR}$--space is exactly $v=g_+$ (just in conformity with the velocity
 value  (C.7) obtainable in case of the planar-type waves).

The {\it
$\cE_g^{SR}$--space extension of the Coulomb law} is given by the formulae:
\be
 B_0=\fr e{\sqrt{\de_{ab}r^ar^b}}=
 \fr{eh}{\xi(g;t)\mt}
=\fr e{j(g;R)\ka(g;R)\mR }
\ee
(the formula (B.38) has been used in the second step and the formula
(B.2) has been applied in the last step);
in the quasi-pseudoeuclidean space,
\be
U_i=k^0_iB_0
\ee
(see (B.38) and (B.39)) with the components
 \be
 U_0=\lf(1+(h-1)\fr{t^0t_0}{S^2}\rg)\fr{e}{\mt},
 \qquad
 U_a=(h-1)\fr{t^0t_a}{S^2}\fr e{\mt};
 \ee
in the $\cE_g^{SR}$--space,
\be
A_p=\rho^0_pB_0
\ee
(see (B.56))
with the components
\be
A_{0}=
\lf(1+\fr1L(h-1+\fr g2\fr{R^0}{\mR}-\fr{g^2}4)\rg)
\fr e{\mR},
\qquad
A_{a}=
-\lf((h-1)\fr{R^0}{\mR}+\fr g2\rg)\fr {R^ae}{\mR^2}.
\ee
In the $O_1(g)$--approximation we get merely
 \be
 U_0=\fr{e}{\mt},
 \qquad
 U_a=0,
 \ee
 and
 \be
A_0=\fr e{\mR}+\fr g2\fr{R^0e}{S^2},
\qquad
A_a=-\fr g2\fr{R^ae}{\mR^2}.
\ee


\ses\ses


\setcounter{sctn}{6}
\setcounter{equation}{0}
{\nin\bf 6. Conformal scalar and spinor fields and solutions}

\ses\ses

In distinction from the electromagnetic field case,
the conventional Lagrangian (2.2)
for a complex scalar field $\phi=\phi(R)$
 is not conformally invariant.
Let us
modify the Lagrangian as follows:
\be
L_{\{\phi\}}=g^{pq}(g;R)\prtl_p \phi^*\prtl_q \phi+\fr1{6}S^{pq}{}_{pq} \phi^*\phi
-\ka^2m^2 \phi^*\phi.
\ee
Since
\be
S^{pq}{}_{pq}
=\fr14g^2\fr 6{F^2(g;R)}
\ee
(see (A.22)--(A.23)), we get
\be
L_{\{\phi\}}=g^{pq}(g;R)\prtl_p \phi^*\prtl_q \phi+\fr14g^2\fr1{F^2} \phi^*\phi
-\ka^2m^2 \phi^*\phi.
\ee
This Lagrangian entails the associated scalar field equation
\be
\fr1{J}\fr{\prtl}{\prtl R^p}
\Bigl(Jg^{pq}\fr{\prtl}{\prtl R^q}\phi\Bigr)
-\fr14g^2\fr1{F^2} \phi+\ka^2m^2 \phi=0
\ee
which extends (2.8)--(2.9).
Using the coordinates
\be
r^n=\rho^n(g;R),
\ee
where  $\rho^n$ are the functions given by (B.44),
let us go over to a new field
\be
a=a(r)
\ee
according to the transformation
\be
\phi(R)=\ka(g;R)a(\rho(g;R))
\ee
($\ka$ is the conformal multiplier (B.43)).
Calculations show that the insertion of (6.7) into (6.4)
results in  the  equation of the ordinary Klein-Gordon type:
\be
e^{ij}
\fr{\prtl^2a}{\prtl r^i\prtl r^j}
+m^2a
=0
\ee
\ses
which is obviously  derivable from the  ordinary pseudoeuclidean Lagrangian
\be
L_{\{a\}}=e^{ij}\fr{\prtl a^*}{\prtl r^i}\fr{ \prtl a}{\prtl r^j}-m^2 a^*a.
\ee
 Accordingly, we introduce

 \ses

 DEFINITION. The field
$
a=a(r)
$
is called the {\it conformal scalar field}.

\ses

In particular,
the plane-wave solutions to the latter field generate the solutions
\be
\phi(R)=\fr1{(2\pi)^{3/2}}\ka(g;R)\int\e^{i\Phi}\delta(k^2-m^2)\phi(k)d^4k,
\ee
where $\Phi$ is the phase (B.55).
By using (2.19),
the latter solutions can be transformed to  the quasi-pseudoeuclidean space:
\be
u(t)=\fr1{(2\pi)^{3/2}}\xi(g;t)\int\e^{i\Phi}\delta(k^2-m^2)u(k)d^4k,
\ee
in which case the representation (B.51) for the phase
$\Phi$ should be taken.


The quasi-pseudoeuclidean versions of the formulae (6.3), (6.4), and (6.7) read
\be
L_{\{u\}}=n^{ij}(g;t)\prtl_i u^*\prtl_j u+\fr14g^2\fr1{S^2} u^*u
-\xi^2m^2u^*u,
\ee
\ses
\be
\prtl_i(n^{ij}u_j)-\fr14g^2\fr1{S^2}u+\xi^2m^2u=0,
\ee
and
\be
u(t)=\xi(g;t)a\Bigl(\fr1h\xi(g;t)t\Bigr),
\ee
where
$u=u(t)=\phi(R)$ in accordance with (2.19) and $\xi$ is the conformal multiplier (B.36).
We get
\be
u_i=\fr{\ga t_i}{S^2}\xi a+\fr1h\xi^2a_i+\fr1h\ga\xi^2
\fr{a_nt^n}{S^2}t_i,
\qquad
t^iu_i=\ga\xi a+\xi^2a_nt^n,
\ee
with
$
a_n(r)= \D{ a(r)}{r^n}$ and $ u_i=\D u{t^i}.
$
Also,
\be
n^{ij}u_j=\ga\xi a\fr{t^i}{S^2}
-\ga\xi^2
\fr{a_nt^n}{S^2}t_i
+k\xi^2a^i,
\ee
where $a^i=e^{ij}a_j$,
\be
\prtl_ja_i=\fr1h\xi(a_{ij}+\ga
\fr{a_{ik}t^k}{S^2}t_j),
\ee
and
\be
a_{ij}(r)=\fr{\prtl^2a(r)}{\prtl r^i\prtl r^j}.
\ee
Finally
\be
\prtl_i(n^{ij}u_j)=\fr{g^2}4\xi\fr a{S^2}+\xi^3\D{a^i}{r^i}.
\ee
Inserting the last result in (6.13) just yields (6.8).

The {\it
$\cE_g^{SR}$--space extension of the Yukawa potential} is given by the following formulae:
\be
a=\fr{\exp\Bigl(-m\sqrt{\de_{ab}r^ar^b}\Bigr)}{m\sqrt{\de_{ab}r^ar^b}}
=
\fr {h\exp\Bigl(-\fr1hm\xi(g;t)\mt\Bigr)}{m\xi(g;t)\mt}
=
\fr{\exp\Bigl(-mj(g;R)\ka(g;R)\mR\Bigr)}{mj(g;R)\ka(g;R)\mR},
\ee
\ses
\be
u(t)=\fr {h}{m\mt}\exp\Bigl(-\fr1hm\xi(g;t)\mt\Bigr)
\ee
(in accordance with the rule (6.14)),
and
\be
\phi(R)=\fr{1}{mj(g;R)\mR}\exp\Bigl(-mj(g;R)\ka(g;R)\mR\Bigr).
\ee
In the $O_1(g)$--approximation these representations reduce to
\be
u(t)=\fr {1}{m\mt}\e^{-m\mt}
\ee
and
\be
\phi(R)=\fr{1}{m\mR}
\lf(
1+\fr14g(1+m\mR)\ln\fr{R^0-\mR}{R^0+\mR}\rg)\e^{-m|{\bf R}|}.
\ee
It is worth comparing (6.20)--(6.24) with (5.61)--(5.67).


For the spinor we take the Lagrangian
\be
L_{\{\psi\}}=\fr i2[\bar\psi\ga^pD_p\psi-(D_p\bar\psi)\ga^p\psi]
-\ka m\bar\psi\psi
\ee
in which the factor $\ka$ in the last term is the only distinction from the standard choice.
This  entails the following generalization of the Dirac equation:
\be
i\ga^pD_p\psi
-\ka m\psi
=0.
\ee
Applying the substitution
\be
\psi(R)=[\ka(g;R)]^{\al}U(\rho(g;R))
\ee
with
\be
\al=\fr32
\ee
reduces the  equation (6.22) to the ordinary pseudoeuclidean form
\be
i\wt\ga^iD_iU- mU=0,
\ee
which can be verified  by straightforward calculations;
the notation
$\wt\ga^i = \ga^Ph_P^i $
has been used.
So,
\be
L_{\{U\}}=\fr i2[\bar U\wt\ga^iD_iU-(D_i\bar U)\wt\ga^iU]
- m\bar UU.
\ee

 Accordingly, we introduce

\ses \ses

 DEFINITION. The field
$
U=U(r)
$
is called the {\it conformal spinor field}.

\ses\ses

In particular, the equation (6.22) admits the wave solution
\be
\psi(R)=\fr1{(2\pi)^{3/2}}[\ka(g;R)]^{\al}\int\e^{i\Phi}\delta(k^2-m^2)\psi(k)d^4k,
\ee
where, as in the ordinary theory,
the amplitude $\psi(k)$ is subjected by the condition
\be
(\ga^Pk_P+m)\psi(k)\Bigl|_{k^2=m^2}=0.
\ee
By means of the transformation (4.11)
the solution can be transformed to  the quasi-pseudoeuclidean space:
\be
v(t)=\fr1{(2\pi)^{3/2}}[\xi(g;t)]^{\al}\int\e^{i\Phi}\delta(k^2-m^2)v(k)d^4k,
\ee
where
$v(k)=\psi(k)$.

The quasi-pseudoeuclidean versions of the formulae (6.21)--(6.23) read
\be
L_{\{v\}}=\fr i2[\bar v\ga^iD_iv-(D_i\bar v)\ga^iv]
-\xi m\bar vv,
\ee
\ses
\be
i\ga^iD_iv
-\xi  m v
=0,
\ee
and
\be
v(t)=[\xi(g;t)]^{\al}U(\fr1h\xi(g;t)t).
\ee
We find
\be
v_n(t)=\ga \al\fr{t_n}{S^2}\xi^{\al} U+\xi^{\al+1}f^m_nU_m,
\ee
\ses
\be
\D U{t^n}=\xi\fr1h(U_n+\fr{\ga}{S^2}t_nt^mU_m)=\xi f^m_nU_m,
\ee
where
$v_n(t)=\prtl v(t)/\prtl t^n$
 and
$U_n(r)=\prtl U(r)/\prtl r^n$,
and also
\be
\ga^nv_n=\ga\al\fr{\ga_jt^j}{S^2}\xi^{\al}U+\xi^{\al+1}\wt\ga^mU_m
\ee
together with
\be
\ga^nD_nv=\ga^n(\prtl_nv+\fr18{R^{PQ}}_n(\ga_P\ga_Q-\ga_Q\ga_P)v)=
\ga(\al-\fr32)\fr{\ga_jt^j}{S^2}\xi^{\al}U+\xi^{\al+1}\wt\ga^mU_m
\ee
(the formulas (4.12) and (4.13) together with (B.35) have been used).
Comparing the last result with (6.24) and (6.31) verifies (6.25).


\ses\ses


\setcounter{sctn}{7}
\setcounter{equation}{0}

\nin {\bf 7. $\cE^{SR}_g$--space quantization of  fields }

\ses\ses


Starting from the conformally flat space referred
 to the coordinates
$r^n
$
given by (6.5),
 the results of the ordinary
relativistic quantum fields  can be used.

We obtain the following wave decompositions:
for the scalar field
\be
\phi(R)
=
\frac1{(2\pi)^{3/2}}
\ka(g;R)\int\e^{ik_nr^n(g;R)}\phi(k)\de(k^2-m^2)d^4k,
\ee
for the electromagnetic vector potential
\be
A_p(R)
=
\frac1{(2\pi)^{3/2}}\rho^j_p(g;R)
\int a_j(k_n)\e^{ik_nr^n(g;R)}\de(k^2)d^4k,
\ee
and for the spinor field
\be
\psi(R)
=
\frac1{(2\pi)^{3/2}}
[\ka(g;R)]^{3/2}\int\e^{ik_nr^n(g;R)}\psi(k)\de(k^2-m^2)d^4k.
\ee

The waves obey the proper value equations in single-wave cases:
\be
\cP_n\phi_{\{k\}}=k_n\phi_{\{k\}},
\qquad
\cP_nA_{\{k\}q}=k_nA_{\{k\}q},
\qquad
\cP_n\psi_{\{k\}}=k_n\psi_{\{k\}},
\ee
where $\cP_n$ is the {\it operator of the four-dimensional momentum} which acts
in the following way:
\be
\cP_n\phi_{\{k\}}=-i\eta^q_n\prtl_q(\fr1{\ka}\phi_{\{k\}})
\ee
 in scalar case,
 \be
\cP_n\psi_{\{k\}}=-i\eta^q_n\prtl_q(\fr1{\ka^{3/2}}\psi_{\{k\}})
\ee
in spinor case,
and
simply
\be
\cP_nA_q=-i\eta^q_n(\prtl_qA_{p}-\eta^s_m\partial_q\rho^m_pA_{s})
\ee
in  electromagnetic  case; the coefficients
$
\rho^m_p
$
and
$
\eta^q_n
$
are given by (B.48) and (B.49).
\ses

The generalized differences
\be
(t\ominus t')^n=\fr1h\xi(g;t)t^n-\fr1h\xi(g;t')t'^n
\ee
(see (B.38)) and
\be
(R\ominus R')^n=\rho^n(g,R)-\rho^n(g;R')
\ee
(see (B.44)) are arisen.


The quasi-pseudoeuclidean versions to the representations (7.1)--(7.7) read:
for the scalar field
\be
u(t)
=
\frac1{(2\pi)^{3/2}}
\xi(g;t)\int\e^{i\xi(g;t) k_nt^n/h}\phi(k)\de(k^2-m^2)d^4k,
\ee
for the electromagnetic vector potential
\be
U_i(t)
=
\frac1{(2\pi)^{3/2}}k^j_i(g;t)
\int a_j(k_n)\e^{i\xi(g;t) k_nt^n/h}\de(k^2)d^4k,
\ee
and for the spinor field
\be
v(t)
=
\frac1{(2\pi)^{3/2}}
[\xi(g;t)]^{3/2}\int\e^{i\xi(g;t) k_nt^n/h}v(k)\de(k^2-m^2)d^4k;
\ee
also,
\be
\cP_nu_{\{k\}}=k_nu_{\{k\}},
\qquad
\cP_nU_{\{k\}j}=k_nU_{\{k\}j},
\qquad
\cP_nv_{\{k\}}=k_nv_{\{k\}},
\ee
with
\be
\cP_nu_{\{k\}}=-ik^j_n\prtl_j(\fr1{\xi}u_{\{k\}}),
\ee
\ses
 \be
\cP_n v\psi_{\{k\}}=-ik^j_n\prtl_j(\fr1{\xi^{3/2}}v_{\{k\}}),
\ee
\ses
and
\be
\cP_nU_j=-ik^j_n(\prtl_jU_i-k^l_m\partial_jK^m_iU_{l}),
\ee
where
$
k^i_j
$
 are the coefficients
(B.39) and
$
K_i^nk^i_j=\de^n_j.
$
\ses


In case of scalar field, the commutator of the operators
$ \phi^-(\bk)$ and $\phi^+(\bq)$ is subjected to  the  condition
 \be
[\phi^-(\bk),\phi^+(\bq)]=\delta(\bk-\bq).
 \ee
In case of the electromagnetic field, it is appropriate to take
the decomposition
 \be
 B_m=e^s_mb^{\pm}_s(\bk)\e^{i\phi}
 \ee
in terms of an orthonormal frame $\{e_i^s\}$, so that
$
e_i^me_j^ke_{mk} =e_{ij},
$
assume the property
\be
b_i^+(\bvec k)b^{i,-}(\bvec k)=e^{mn}a^+_m(\bvec k)a^-_n(\bvec k),
\ee
and set forth the commutator   condition
 \be
[a^-_i(\bvec k),a^+_j(\bvec q)]=-e_{ij}\delta(\bvec k-\bvec q).
 \ee
To quantize the Fermion field, we regard the amplitudes
 $a_s(\bk)$  as some operators which satisfy the permutation relations of Fermi-Dirac
 type:
\be
\{(a^-_s(\bk))^*,a^+_r(\bq)\}=\{a^-_s(\bk),(a^+_r(\bq))^*\}=
\delta_{sr}\delta(\bk-\bq).
\ee


So, the required commutators for fields in the $\cE_g^{SR}$--space under consideration
read as follows:
in scalar case,
 \be
[\phi^-(R),\phi^+(R')]=  \fr1i
\ka(g;R)\ka(g;R')
\De^-(R\ominus R'),
\ee
\ses
 \be
[\phi^+(R),\phi^-(R')]= \ka(g;R)\ka(g;R')
\fr1i\De^+(R\ominus R'),
\ee
and
 \be
[\phi(R),\phi(R')]= \ka(g;R)\ka(g;R')
\fr1i\De(R\ominus R');
\ee
in electromagnetic case,
\be
[A^-_p(R),A^+_q(R')]
=is_{pq}(g;R,R')\De_0^-(R\ominus R'),
 \ee
\ses
 \be
[A^+_p(R),A^-_q(R')]=is_{pq}(g;R,R')\De_0^+(R\ominus R'),
 \ee
and
\be
[A_p(R),A_q(R')]=is_{pq}(g;R,R')\De_0(R\ominus R'),
\ee
where the tensor
\be
s_{pq}(g;R,R') =\rho^m_p(g;R)\rho^k_q(g;R')e_{mk}
\ee
appears;
in spinor case,
\be
[\psi(R),\bar\psi(R')]=\frac1i
[\ka(g;R)\ka(g;R')]^{3/2}
S(R\ominus R').
\ee


Converting the consideration into the quasi-pseudoeuclidean space,
we obtain the following result for the respective commutators:
for the scalar field,
 \be
[u^-(t),u^+(t')]=\xi(g;t)\xi(g;t') \fr1i\De^-(t\ominus t'),
\ee
\ses
 \be
[u^+(t),u^-(t')]= \xi(g;t)\xi(g;t')
\fr1i\De^+(t\ominus t'),
\ee
and
 \be
[u(t),u(t')]= \xi(g;t)\xi(g;t')
\fr1i\De(t\ominus t');
\ee
for the electromagnetic field
\be
[U^-_i(t),U^+_j(t')]
=
ic_{ij}(g;t,t')\De_0^-(t\ominus t'),
 \ee
\ses
  \be
[U^+_i(t),U^-_j(t')]=ic_{ij}(g;t,t')\De_0^+(t\ominus t'),
 \ee
and
\be
[U_i(t),U_j(t')]=ic_{ij}(g;t,t')\De_0(t\ominus t'),
\ee
where
\be
c_{ij}(g;t,t')=k_i^m(g;t)k_j^k(g;t')e_{mk};
\ee
for the spinor field
\be
[v(t),\bar v(t')]=\frac1i
[\xi(g;t)\xi(g;t')]^{3/2}
S(t\ominus t').
\ee


In the very conformally flat space,
the ordinary pseudoeuclidean relations take place, so that
in scalar case,
$$
[\phi^-(r),\phi^+(r')]=  \fr1i\De^-(r - r'),
$$
\ses
$$
[\phi^+(r),\phi^-(r')]= \fr1i\De^+(r - r'),
$$
\ses
$$
[\phi(r),\phi(r')]= \fr1i\De(r - r');
$$
in electromagnetic case,
$$
[B^-_i(r),B^+_j(r')]
=ie_{ij}\De_0^-(r - r'),
$$
\ses
$$
[B^+_i(r),B^-_j(r')]=ie_{ij}\De_0^+(r - r'),
$$
\ses
$$
[B_i(r),B_j(r')]=ie_{ij}\De_0(r-r');
$$
and in spinor case,
$$
[U(r),\bar U(r')]=\frac1iS(r-r');
$$
with the permutation functions
$$
\De^+(r)=\fr1{(2\pi)^3i}\int\theta(k_0)e^{ik_nr^n}\delta(k^2-m^2)d^4k,
$$
\ses
$$
\De^-(r)=\fr i{(2\pi)^3}\int\theta(-k_0)e^{-ik_nr^n}\delta(k^2-m^2)d^4k=
-\De^+(r),
$$
\ses
$$
\De(r)=\De^+(r)+\De^-(r)
= \fr i{(2\pi)^3}\int\epsilon(k_0)e^{-ik_nr^n}\delta(k^2-m^2)d^4k,
$$
\ses
$$
\De_0^+(r)=\fr1{(2\pi)^3i}\int\theta(k_0)e^{ik_nr^n}\delta(k^2)d^4k,
$$
\ses
$$
\De_0^-(r)=\fr i{(2\pi)^3}\int\theta(-k_0)e^{-ik_nr^n}\delta(k^2)d^4k=
-\De_0^+(r),
$$
\ses
$$
\De_0(r)=\De_0^+(r)+\De_0^-(r)
= \fr i{(2\pi)^3}\int\epsilon(k_0)e^{-ik_nr^n}\delta(k^2)d^4k,
$$
and
$$
S(r)=(i\gamma^j\fr{\partial}{\partial r^j}+m)\De(r).
$$



\ses\ses

\setcounter{sctn}{8}
\setcounter{equation}{0}
{\nin\bf 8.
$\hat\cE_g^{SR}$--way regularization
 }

\ses\ses

The presence of the geometrical parameter  $g$ can be used
for constructing an appropriate function
$\cW$ which ought to play the role of
the {\it  $\hat\cE_g^{SR}$--space weight} when performing integration
over the
four-dimensional momenta~ $\{P_q\}\in\hat V_4$.
\ses

{\it Suppressing ultraviolet divergences} can be motivated, first of all, by the circumstance
that,
since the  $\hat\cE_g^{SR}$--space approach under study is based upon a
group
of invariance (which transformations leave
the FHF $H$ in the co-space
$\hat V_4$ invariant),
the function $\cW$
should include the dependence on  $\{P_q\}$ through the modulus
$P=H(g;P_q)$.
Next, when integrating over
 $\{P_q\}$,
it is natural to choose the function $\cW$ to be a decreasing exponent:
\be
\cW(g;P) = C_1(g)
e^{-\al(g)P^{\nu}},
\ee
where $\al(g)>0$, $\nu>0$,  and $C_1(g)$ stands for a positive normalizing factor, so that
\be
\int_{P=0}^\iy
\cW(g;P)
P^3
dP
=1
\ee
and
\be
{ C_1(g)}=
\fr1{\int_{P=0}^\iy
\cW(g;P)
P^3
dP}.
\ee
The answer to the question ``what choice
is to be preferred for the factor
$\al(g)$''
can  tentatively  be given by realizing
the attractive idea of geometrical origin that the factor
stems from the
  $\hat\cE_g^{SR}$--space
curvature. For the later  is proportional to
~$g^2$
 (see (A.22) and (A.23)), we suggest the choice
 \be
 \al = Cg^2,
 \ee
where $C$ is a positive  constant.
The respective
invariant
integration with the   $\hat\cE_g^{SR}$--space  weight
$\cW$ over the space
$\hat V_4$ should be performed in accordance with the rule
\be
\int_{\hat V_4}\cM = \int_{P=0}^\iy Z(g;P)\cW(g;P)P^3dP, \qquad
Z = \int \cM \hat Jd\hat\Omega ,
\ee
where $\cM$ is some matrix element or scalar,
$\hat J=\sqrt{|\det(g_{pq}(g;P_s))|}$,
and $d\hat\Omega$ is an appropriate angular element of volume.
Because of the exponential form  (8.1) of the weight  $\cW$, the
integral with respect to $P$ will be converged at the higher limit for any degrees
 $z$ of the terms proportional to $P^z$.
Therefore, upon the Finslerian extension under consideration,
the finite expressions proportional to natural degrees of
the square of the inversed   $\hat\cE_g^{SR}$--space
parameter, $1/g^2$,
will relate to the divergent integrals that appear in ordinary theory at
 $P\to\iy$.
In this sense the   $\hat\cE_g^{SR}$--space  approach developed
may give, in principle,  a simple explanation for the nature of the ultraviolet divergences
(which is characteristic for the ordinary
Lorentz-invariant theory of quantized fields:
the limiting transition $g\to0$ in the factor (8.3)
is far from always being possible!
Though in the exponent of the   $\hat\cE_g^{SR}$--space  weight~(8.1)
it is possible to go over to the limit
$g\to0$,
which does remove the exponent (for $e^{-Cg^2P^{\nu}}\to1$ at $g\to0$),
the normalizing factor $C_1(g)$ cannot be analytical
at $g=0$.
For example,
it is  worth choosing
\be
\cW(g;P) = C_1(g)
e^{-g^2P^{2}/m^2},
\ee
where $m\ne0$ is the mass of particle,
and
use the Fourier transformation
\be
\fr1{(2\pi)^4}\int\e^{ik_nr^n-g^2(k^2/m^2)}h^{-2}d^4k=
\fr1{16\pi^2h^2}\fr{m^4}{g^4}\e^{-r^2m^2/4g^2}.
\ee
In this way, the so-called alpha-representation
\be
\fr1{m^2-k^2-i\epsilon}=i\int\e^{i\al(k^2-m^2+i\epsilon)}d\al
\ee
can be applied to find form-factors and evaluate various diagrams and matrix elements.


\ses

{\it  Suppressing angular divergences},
which are stemming from integrations over the particle energy variable $P_0$,
is of an alternative need.
 By following the known general ideas and methods of statistical physics, we can
take the respective weight
$\cW_1$ in the form of the exponent
\be
\cW_1= C_3(g) e^{-C_2g^2P_0},
\ee
where $C_2$ and $C_3$ are positive constants and
$P_0$
is the particle energy.
To have in the exponent the energy in dimensionless combination,
we may substitute    the normalized energy
\be
K_0=P_0/m
\ee
with  $P_0$.
In the on-mass-shell treatment, that is, when one sets forth
the dispersion relation
\be
H(g;P_q)=m
\ee
(with the FHF $H$ given by (A.14)),
the phase volume  element is to be taken as
\be
d\hat V\eqdef\delta\lf((H(g;P_q))^2-m^2\rg)\hat Jd^4P
=
\hat J
\fr{
d^3
{\bf P}
}{2P^0}\Bigl.\Bigr|_{H(g;P_q)=m}
\ee
and the weight takes on the form
\be
\cW_1(g,|{\bf P}|) = C_3(g)e^{-C_4g^2K_0(g;|{\bf P}|/m)}
\ee
with the function
$
K_0=K_0(g;|{\bf P}|/m)
$
obtainable from the equation $H(g;K_0,{\bf P}/m)=1$.
Because of the exponential
form (8.13)
 the integral over
$|{\bf P}|$
will converge at the higher limit at any degrees
$|{\bf P}|^n$
in the matrix element, so that the divergent integrals appeared in the ordinary
pseudo-euclidean theory at the process
 $|{\bf P}|\to\iy$,
upon   $\cE_g^{SR}$--space  extension can be juxtaposed with finite expressions
proportional to degrees of
$1/g^2$.
On neglecting the parameter $g$ in the function
$
K_0=K_0(g;|{\bf P}|/m),
$
the weight function (8.12) is simplified to read
\be
\cW_1(g,|{\bf P}|) = C_3(g)e^{-C_4g^2\sqrt{1+(|{\bf P}|/m)^2}}.
\ee
In such a case, the normalization
\be
\int
\cW_1(g,\mP)
d^3
{\bf P}=1
\ee
entails
\be
\fr1{4\pi C_3(g)}=2\lf(\fr8{g^2}\rg)^2K_1\lf(\fr{g^2}8\rg)
+
\lf(\fr8{g^2}\rg)
K_0
\lf(\fr{g^2}8\rg),
\ee
\ses\\
where $\acute K_0$ and $\acute K_1$ ---
the Macdonald functions (the Hankel functions of imaginary argument), for which
\be
\int_0^\iy e^{-z\ch u}\sh^2u\, du=\fr1z\acute K_1(z)
\ee
and
\ses
\be
\acute K'_1(z)=-\fr1z\acute K_1(z)-\acute K_0(z).
\ee
In the law-energy  region
\be
\fr{|{\bf P}| }{m}\ll 1
\ee
we get the Maxwell-type distribution
\be
\cW_1(g,|{\bf P}|)
= C_3(g)e^{-C_4g^2|{\bf P}|^2/2{m}^2}.
\ee
Alternatively, in the ultrarelativistic region
\be
\fr{|{\bf P}| }{m}\gg 1
\ee
we get the approximation
\be
\cW_1(g,|{\bf P}|)
= C_3(g)e^{-C_4g^2|{\bf P}|/{m}}.
\ee






\setcounter{sctn}{1} \setcounter{equation}{0}

{\nin\bf Appendix A. Basic properties of the space $\cE_g^{SR}$}

\ses \ses

Searching for extension of the pseudoeuclidean geometry
in due Finsler-relativistic  way,
we should
adapt constructions to the following decomposition
of the four-dimensional vector space $V_4$:
 \be
V_4=\cS_g^+\cup \Si_g^+\cup{\cR_g}\cup\Si_g^-\cup\cS_g^-,
\ee
which sectors relate to the cases when the  contravariant vector
 $R\in V_4$ is respectively
 future--timelike, future--isotropic, spacelike, past--isotropic, and
 past--timelike.
The respective co-analogue for the covariant vectors (the momenta) $P\in\hat V_4$
reads
 \be
  \hat V_4=
\hat\cS_g^+\cup\hat\Si_g^+\cup\hat\cR_g\cup\hat\Si_g^-\cup\hat\cS_g^-.
\ee

With this purpose, we introduce the following convenient notation:
\be G= g/h, \ee
\medskip
\be \qquad h = \sqrt{1+\fr14g^2}, \ee
\ses
\be g_+=-\fr12g+h,
\qquad g_-=-\fr12g-h,
\ee
\medskip
\be G_+=g_+/h\equiv -\fr12G+1, \qquad G_-=g_-/h\equiv -\fr12G-1,
\ee
\medskip
\be
 g^+=1/g_+=-g_-,  \qquad  g^-=1/g_-=-g_+,
 \ee
\medskip
\be
 g^+=\fr12g+h, \qquad g^-=\fr12g-h,
 \ee
\medskip
\be
 G^+=g^+/h\equiv \fr12G+1, \qquad G^-=g^-/h\equiv \fr12G-1.
 \ee

We shall decompose vectors to select the timelike components and the
three-dimensional spatial components:
 $ R=\{R^0,\bR\}$ and $ P=\{P_0,\bP\}. $
In terms of the forms \be B(g;R)=-\lf(R^0+g_-|{\bf
R}|\rg)\lf(R^0+g_+|{\bf R}|\rg)
\equiv-\lf((R^0)^2-gR^0\mR-\mR^2\rg)
 \ee
and
\be
 \hat B(g;P)=-\lf(P_0-\fr{|{\bf P}|}{g^+}\rg)\lf(P_0-\fr{|{\bf
P}|}{g^-}\rg) \equiv-\lf((P_0)^2+gP_0\mP-\mP^2\rg), \ee
\ses\\
all the sectors that enter the decompositions (A.1) and (A.2)
can be embraced by one FMF
 \be
F(g;R)=\sqrt{|B(g;R)|}\,j(g;R)=\lf|R^0+g_-|{\bf
R}|\rg|^{G_+/2}\lf|R^0+g_+|{\bf R}|\rg|^{-G_-/2}, \ee where \be
j(g;R)=\lf|\fr{R^0+g_-|{\bf R}|}{R^0+g_+|{\bf R}|}\rg|^{-G/4}, \ee
\ses\\
and one FHF \be H(g;P)=\sqrt{|\hat
B(g;P)|}\,\hat j(g;P)=\lf|P_0-\fr{|{\bf P}|}{g^+}\rg|^{G^+/2}
\lf|P_0-\fr{|{\bf P}|}{g^-}\rg|^{-G^-/2}, \ee where \be \hat
j(g;P)=\lf|\fr{P_0-\fr{|{\bf P}|}{g^+}}{P_0-\fr{|{\bf
P}|}{g^-}}\rg|^{G/4}. \ee

\ses

By following the methods of the Finsler geometry,
we use the definitions for the covariant vector
$$
R_p \eqdef \fr12\D{F^2(g;R)}{R^p} \equiv P_p
$$
and the FMT
$$
g_{pq}(g;R) \eqdef \fr12\, \fr{\prtl^2F^2(g;R)}{\prtl R^p\prtl
R^q} =\fr{\prtl R_p(g;R)}{\prtl R^q};
$$
$p,q,...=0,1,2,3$.

Thus we get the {\it pseudo-Finsleroid relativistic space
 }
  \be
\cE_g^{SR}=\{V_4;\,F(g;R);\,g_{pq}(g;R);\,R\in V_4\}
 \ee
and the
{\it pseudo-Finsleroid relativistic co-space }
 \be
\hat\cE_g^{SR}=\{\hat V_4;\,H(g;P);\,g^{pq}(g;P);\,P\in\hat V_4\}.
\ee



Special calculations can be used to verify the equalities
\ses\\
\be \det(g_{pq}(g;R))=-[j(g;R)]^8 \ee
and
\be
\sign(g_{pq})=\sign(g^{pq})=(+ - - -). \ee

The following assertion is valid:
 \it for the pseudo-Finsleroid space $\cE_g^{SR}$ the Cartan torsion tensor
$$
C_{pqr} \eqdef \fr12\D{g_{pq}}{R^r}
$$
is of the special algebraic form
 \rm
\be
C_{pqr}=\fr1N\lf(h_{pq}C_r+h_{pr}C_q+h_{qr}C_p-\fr1{C_tC^t}C_pC_qC_r\rg),
\ee \it
where
\be C_pC^p=-\fr{N^2g^2}{4F^2} \ee
and $N=4$. \rm
Here, $C_p=\3Cpqq$.

Proof is gained by straightforward calculations
on the basis of the explicit form of components of the FMT
and the Cartan tensor (see more detail in  [9--11]). Inserting
(A.20)--(A.21) in the general expression for the curvature tensor
$$
S_{pqrs} = C_{tqr}\3Cpts-C_{tqs}\3Cptr
$$
yields the following simple result after rather simple straightforward calculations:
\be S_{pqrs}=S^*
(h_{pr}h_{qs}-h_{ps}h_{qr})/F^2
 \ee
with the constant
 \be S^*=\fr14g^2.
\ee \rm \ses
The tensor \be h_{pr} \eqdef g_{pr} -l_pl_r
\ee
has been used, where  $ l_p=R_p/F(g;R)$ --- the unit vector components.


The FMF  (A.12) defines the {\it pseudo-Finsleroid}
 \be
  \cF^{\text{Relativistic}}_g ~:=\{R\in  V_4:F(g;R)\le 1\}.
   \ee
The associated
{\it
 indicatrix $\cI_g $} defined by
 \be
 \cI_g :=\{R\in V_4:F(g;R)=1\}
  \ee
is the surface of the pseudo-Finsleroid.
With the given FHF (A.14), the body
 \be
\hat\cF^{\text{Relativistic}}_g := \{\hat R\in \hat V_4: H(g;\hat
R)\le1\}
\ee
 is called the \it co-pseudo-Finsleroid\rm.
The respective figuratrix introduced according to
 \be
  \hat\cI_g := \{\hat R\in \hat V_4: H(g;\hat
R)=1\}
 \ee
is called the  \it co-indicatrix\rm.

\ses

From (A.22)--(A.24) it follows that   \it in case of the pseudo-Finsleroid space
 $\cE_g^{SR}$ the indicatrix is a space of the constant negative curvature
 which value is equal to
\rm \be
R_{\text{Ind}}=
-\lf(1+\fr14g^2\rg)\le-1.
 \ee
\ses

The respective    {\it equation of  $\cE_g^{SR}$--geodesics} is of the form
\be
 \fr{d^2R^p}{ds^2}+\3Cqpr(g;R)\fr{dR^q}{ds}\fr{dR^r}{ds}=0,
 \ee
where $s$ is the parameter of the arc-length defined in accordance with
the rule
 \be
  ds
\eqdef \sqrt{| g_{pq}(g;R)dR^pdR^q|}.
 \ee

The use of the functions
\be
A(g;R)=R^0-\fr12g\mR,\qquad \hat A(g;P)=P_0+\fr12g\mP \ee
and
\be
L(g;R)=\mR-\fr12gR^0,\qquad \hat L(g;P)=\mP +\fr12gP_0
\ee
is often convenient in various calculations.

In the limit  $g\to 0$
the considered space degenerates to the ordinary pseudoeuclidean case:
$$
 B|_{_{g=0}}= -[(R^0)^2-{\bf R}^2],
\qquad \hat B|_{_{g=0}}= -[(P_0)^2-{\bf P}^2],
$$
\ses
$$
j|_{_{g=0}}= \hat j|_{_{g=0}}= 1, \qquad F|_{_{g=0}}=
\sqrt{|(R^0)^2-{\bf R}^2|},
$$
\ses
$$
H|_{_{g=0}}= \sqrt{|(P_0)^2-{\bf
P}^2|},
 \qquad
g_{pq}|_{_{g=0}}= e_{pq},
$$
\ses
$$
g^{pq}|_{_{g=0}}= e^{pq}, \qquad C_{pqr}|_{_{g=0}}= 0, \qquad R_{\rm
Ind}|_{_{g=0}}= -1.
 $$
 Since at $g=0$ the space $\cE_g^{SR}$
is pseudoeuclidean,
then
 $
\cI_{g=0}
 $
is the ordinary unit hyperboloid.


The
{\it
 $\cE_g^{RS}$--space invariant orthonormal frames:
}
\be
e^P_q~:= \si^r_qf^P_r,\qquad
e_P^q~:= \mu_r^qm^r_P
\ee
are constructed on the basis of the formulas
(B.27)--(B.31).
 The reciprocity
 $e^P_qe^q_R=\de^P_R$
holds.
Calculations show that
\be
e^P_q=\fr1h\si^P_q+\fr{h-1}h\si^Pl_q/F,\qquad
e_P^q=h\mu_P^q+(1-h)\si_Pl^q/F,
\ee
and
\be
g_{pq}(g;R)=e_p^{0}(g;R)e_q^{0}(g;R)-
\suml_{c=1}^{3}e_p^{c}(g;R)e_q^{c}(g;R).
\ee
The
{\it
$\cE_g^{RS}$--space Ricci rotation coefficients}
\be
{R^{PQ}}_p(g;t)~:=\lf(\prtl_ie_j^Q-\3Cprs e_r^Q\rg)e_T^se^{TP}
\ee
can be obtained as the transform
\be
{R^{PQ}}_p(g;R)
=
{R^{PQ}}_i(g;t)\si^i_p
\ee
from the quasi-pseudoeuclidean space.
After using the formula (B.33) and (B.34), we find  explicitly the components
\be
{R^{PQ}}_q(g;R)
=
(h-1)(\si^Pe_q^Q-\si^Qe_q^P)/F^2(g;R),
\ee
or
in another form
\be
{R^{PQ}}_q(g;R)
=
\fr{h-1}h(\si^P\si_q^Q-\si^Q\si_q^P)/F^2(g;R).
\ee
\ses
We have
\be
\si^P/K=L^P,\qquad
e^q_PL^P=l^q,\qquad
e^q_P\si^P=R^q,\qquad
{R^{PQ}}_pR^p=0,
\ee
and
\be
e_{Pt}e_{Qs}(
\partial_p
{R^{PQ}}_q
-
\partial_q
{R^{PQ}}_p)
=\fr{2(h-1)h}{F^2}(
h_{pt}h_{qs}
-
h_{ps}h_{qt}
)
=\fr{2h}{1+h}S_{pqts},
\ee
where
$
e_{Pt}=g_{rt}e_P^r
$.
Also,
$$
\partial_p
{R^{PQ}}_q
-
\partial_q
{R^{PQ}}_p
=
$$
\ses
\be
\fr{2(h-1)h}{F^2}
\Bigl[
e^P_pe^Q_q
-
e^Q_pe^P_q
+\fr1F(
-
e^P_p\si^Ql_q
-
e^Q_q\si^Pl_p
+
e^Q_p\si^Pl_q
+
e^P_q\si^Ql_p
)
\Bigr]
\ee
\ses
and
\be
R_{rsp}=(h-1)(l_rh_{sp}-l_sh_{rp})/F.
\ee

It will be noted that the frames (A.34)
 don't obey the
 ``kinematic"
  property $e_0^p\parallel R^p$. Instead, we have
\be
e_q^P|_{_{g=0}}= \de_q^P,
\qquad
e_P^q|_{_{g=0}}= \de_P^q.
\ee


\ses\ses

\setcounter{sctn}{2} \setcounter{equation}{0}

{\nin\bf Appendix B.  Quasi--pseudoeuclidean transformation}

\ses\ses

Let us introduce the nonlinear transformation
\be t^i=\si^i(g;R) \ee \ses
with the functions \ses \be \si^0=
\lf|\fr{R^0+g_-|{\bf R}|}{R^0+g_+|{\bf R}|}\rg|^{-G/4}
\lf(R^0-\fr12g\mR\rg), \qquad \si^a= h\lf|\fr{R^0+g_-|{\bf
R}|}{R^0+g_+|{\bf R}|}\rg|^{-G/4} R^a; \ee
$i,j,...=0,1,2,3 $ and
$a,b,...=1,2,3$.
With the help of the transformation, we can go over  from the  variables
$\{R^p\}$ to the new variables $\{t^i\}$. The inverse transformation
 \be
  R^p=
\mu^p(g;t)
 \ee
involves the functions
\be
 \mu^0=
\lf|\fr{t^0-m}{t^0+m}\rg|^{G/4} (t^0+\fr12Gm), \qquad \mu^a= \fr1h
\lf|\fr{t^0-m}{t^0+m}\rg|^{G/4} t^a, \ee
\ses\\
so that
 \be t^i\equiv\si^i\lf(g; \mu(g;t)\rg), \qquad
R^p\equiv\mu^p\lf(g; \si(g;R)\rg). \ee The notation
$$
m=\sqrt{|e_{ab}t^at^b|}\in[0,\iy)
$$
has been used; the constant $h$ is given by the formula (A.4).

Let us introduce the pseudoeuclidean metric function
 \be
S(t)\eqdef\sqrt{|e_{ij}t^it^j|} \equiv \sqrt{|(t^0)^2-m^2|} \ee
 $
(e_{ij}={\rm diag}(1,-1,-1,-1) $ --- the pseudoeuclidean metric tensor; $e^{ij}=e_{ij}$).
It can readily be verified that the insertion of the  functions
 (B.2) in (B.6)
yields the identity
\be F(g;R)=S(t) \ee
with the function $F(g;R)$ which is exactly the FMF
 (A.12). In this way, we call
 (B.1)--(B.2)
the  \it quasi-pseudoeuclidean transformation\rm.


\ses

The functions (B.2) are obviously homogeneous of degree 1
with respect to the variable ~$R$:
 \be
\si^i(g;bR)=b\si^i(g;R), \qquad b>0,
 \ee
from which it ensues that the derivatives
 \be
 t_p^i(g;R)\eqdef\D{\si^i(g;R)}{R^p} \equiv
\si_p^i(g;R)
 \ee
obey the identity
 \be
  t_p^i(g;R)R^p=t^i.
 \ee Calculating the determinant gives merely
 \be \det(t_p^i)=j^4h^3.
  \ee

Similarly,
\be
 \mu^p(g;bt)=b\mu^p(g;t), \qquad b>0,
 \ee
 \ses
 \be
\mu_i^p(g;t)\eqdef\D{\mu^p(g;t)}{t^i}\equiv R^p_i,
 \ee
  and
\be
\mu_i^p(g;t)t^i=R^p.
 \ee

Next, let us now construct the tensor
 \be n^{ij}(g;t)\eqdef t_p^it_q^jg^{pq}.
  \ee
Straightforward rather lengthy calculations
result in the following simple representations
 \be
  n^{ij}(g;t)=h^2e^{ij}-\fr14g^2l^il^j,
\qquad n_{ij}(g;t)=\fr1{h^2}e_{ij}+\fr14G^2l_il_j
\ee
\ses\\
($e_{ij}e^{jm}=\de_i^m$ and $n_{ij}n^{jm}=\de_i^m$),
where
 \be
l^i\eqdef t^i/S(t), \qquad l_i\eqdef e_{ij}l^j \ee
are respective pseudoeuclidean unit vectors.
For them the equalities
$$
l^il_i=1, \quad n_{ij}l^j=l_i, \quad n^{ij}l_j=l^i, \quad
n_{ij}l^il^j=1, \quad n_{ij}t^it^j=S^2
$$
are valid.
The inversion of (B.15) can be written in the form
\ses\\
\be g_{pq}(g;R)=n_{ij}\lf(g;\si(g;R)\rg)t_p^i(g;R)t_q^j(g;R). \ee
\ses\\
We also obtain \be \det(n_{ij})=-h^6. \ee


\ses

We call the tensor $\{n_{ij}\}$ with the components (B.16)
 {\it quasi-pseudoeuclidean metric tensor},
and the very space \be \cK_g := \{V_4;\,S(t);\,n_{ij}(g;t);\,t\in
V_4\}
 \ee
  \it quasi-pseudoeuclidean space. \rm
   The formulas (B.7) and (B.15)
show explicitly that  the space defined is
quasi-pseudoeuclidean image of the pseudo-Finsleroid relativistic
space $\cE_g^{SR}$, such that   {\it when using
the quasi-pseudoeuclidean transformations the studied pseudo-Finsleroid relativistic space
 $\cE_g^{SR}$
transforms in the quasi-pseudoeuclidean space
 $ \cK_g=\si(\cF_g)$  differed essentially from the pseudoeuclidean space
 $\cE\equiv\cK_{g=0}$. }

Let us evaluate from the tensor (B.16) the associated Christoffel symbols
 \be
\3Nmin=n^{mk}N_{ikj}, \qquad N_{ikj}=\fr12(n_{ik,j}+n_{jk,i}-n_{ij,k}).
\ee
We have subsequently
 \be
n_{ik,j}\eqdef\D{n_{ik}}{t^j}=\fr14g^2(H_{ij}L_k+H_{kj}L_i)/S,
\ee
\ses \be H_{mi}=e_{mi}-L_mL_i \equiv h^2(n_{mi}-L_mL_i), \ee \ses
\be L^iH_{ij}=0, \ee \ses \be N_{mjn}=\fr14G^2H_{mn}L_j/S, \qquad
\3Nmin =\fr14G^2H_{mn}L^i/S, \ee
and
 \be
\3Nimj(t)=\fr14G^2L^mH_{ij}/S. \ee
This entails the properties
$$ t^i\3Nimj=0, \qquad \3Nijj=0. $$



The {\it quasi-pseudoeuclidean orthonormal frames}
$\{m^i_P\}$ and $\{f^P_i\}$
subjected to the conditions
\be
n^{ij}=e^{PQ}m_P^im_Q^j, \qquad
n_{ij}=e_{PQ}f^P_if^Q_j,
\ee
prove to be taken in the simple form
\ses\\
\be
f_i^P(g;t)=\fr1hh_j^P+\fr{h-1}hL_jL^P
\ee
and
\be
m_P^j(g;t)=hh_P^j+(1-h)L_PL^j
\ee
with the vectors
\be
L^P=h_j^PL^j, \qquad L_P=h_P^jL_j.
\ee
Here,
$\{h^P_j\}$ and $\{h_P^j\}$
--- orthonormal frames for
the pseudoeuclidean metric tensor, so that
\be
e^{ij}=e^{PQ}h_P^ih_Q^j,\qquad
e_{ij}=e_{PQ}h^P_ih^Q_j.
\ee
The identity
\be
f_i^P(g;t)t^i=t^P\equiv h^P_jt^j
\ee
holds.

The  {\it associated quasi-pseudoeuclidean Ricci rotation coefficients}
\be
{R^{PQ}}_i(g;t)~:=\lf(\prtl_if_j^Q-\3Ninjf_n^Q\rg)m_T^je^{TP}
\ee
are found merely as
\be
{R^{PQ}}_i(g;t)=(h-1)(L^Pf_i^Q-L^Qf_i^P)/S(t).
\ee

Also,
\be
\fr18\ga^i{R^{PQ}}_i(\ga_P\ga_Q-\ga_Q\ga_P)
=\fr14\ga^i{R^{PQ}}_i\ga_P\ga_Q=-\fr32(h-1)\fr{\ga_jt^j}{S^2}.
\ee



\ses\ses

{ PROPOSITION}.\it  The quasi-pseudoeuclidean metric tensor
$\{n_{ij}\}$
is {\it conformal} to  the pseudoeuclidean metric tensor.
The conformal multiplier is equal to
$\xi^{-2}$ with
\be
\xi(g;t)=\lf[\frac12S^2(t)\rg]^{\ga/2},
\ee
where
\be
\ga=h-1.
\ee
\rm

\ses\ses

Indeed, applying the transformation
\be
r^i=\xi(g;t)t^i/h
\ee
and using the coefficients
\be
k_j^i~:= \D{r^i}{t^j}
=(\xi\de_j^i+\xi't^it_j)/h
=\xi\lf(\de^i_j+\ga\fr{t^it_j}{S^2(t)}\rg)/h
\ee
to define the tensor
\be
c^{ij}(g;r)~:= k_n^ik_m^jn^{nm}(g;t),
\ee
we obtain merely
\be
c^{ij}=\xi^2e^{ij}.
\ee




Therefore, the tensor
\be
s_{pq}(g;R)=[\ka(g;R)]^2g_{pq}(g;R)
\ee
meaningful over the initial $\cE^{SR}_g$--space is conformally flat, where
\be
\ka(g;R)=\lf[\frac12F^2(g;R)\rg]^{\ga/2}\equiv\xi(g;t).
\ee
The functions
\be
\rho^i(g;R)=\fr1h\si^i(g;R)\ka(g;R)\equiv r^i
\ee
realize the transition from the initial FMT $\{g_{pq}\}$ to the pseudoeuclidean tensor
$\{e_{ij}\}$.
The inverse functions will be denoted as
\be
\eta^p=\eta^p(g;\rho)
\ee
so that
\be
\eta^p(g;\rho(g;R))=R^p.
\ee
This entails the identities
\be
\eta^p_m\rho^m_q=\de^p_q,
\qquad
\eta^p_m\rho^n_p=\de^n_m,
\ee
for the derivatives
\be
\rho^m_p=\D{\rho^m}{ R^p}
\ee
and
\be
\eta_m^p=\D{\eta^p}{ \rho^m}.
\ee
\be
s_{pq}(g;R) =\rho^m_p(g;R)\rho^k_q(g;R)e_{mk}
\ee


The  $\cE^{SR}_g$--space {\it wave phase}
\be
\Phi~:  =\fr1hk_nt^n\xi
\ee
 obeys the important identity
\be
\D{\Phi}{t^j}=\xi f^P_jk_p
\ee
which entails directly that
\be
n^{ij}
\fr{\prtl \Phi}{\prtl t^i}
\fr{\prtl \Phi}{\prtl t^j}
=\xi^2e^{ij}k_ik_j
\ee
(see (B.27)). Inserting the quasi-pseudoeuclidean transformations  (B.2)
 in
(B.35) yields the representation
\be
\Phi=k_n\rho^n(g;R),
\ee
so that
\be
\Phi(g;R^0,R^1,R^2,R^3)=j\ka(\fr1hAk_0-k_aR^a).
\ee
\ses

At the $O_1(g)$--level of consideration the phase (B.55)  reduces to read
\be
\Phi\approx
\lf(1-\fr14g\ln\fr{R^0-\mR}{R^0+\mR}\rg)k_pR^p
-\fr12g\mR k_0,
\ee
where
$
S^2=(R^0)^2-\mR^2
$.



\ses\ses

\setcounter{sctn}{3} \setcounter{equation}{0}

{\nin\bf Appendix C. Zero-phase conformally flat fronts in  $\cE^{SR}_g$--space}

\ses\ses

The zero-phase is described by the equation
\be
\Phi(g;R^0,R^1,R^2,R^3)=0
\ee
which, in view of (B.55), can explicitly be written as
\be
(R^0-\fr12g\mR)k_0-hk_aR^a =0.
\ee
The last equation determines the {\it zero-phase wave surface,}
which is obviously non-plane unless $g=0$.
If for definiteness we specify the components $\{k_0,k_1,k_2,k_3\}$
 to have $k_2=k_3=0$ in accordance with (5.13), the equation (C.2) reduces to merely
\be
R^0-\fr12g\mR-hR^1 =0,
\ee
or explicitly,
\be
R^0=hR^1+\fr12g\mR.
\ee
Therefore, the point of intersection of the wave with the $R^1$--axis
\be
R^1>0, \quad R^2=R^3=0
\ee
is governing by the equation
\be
R^1=g_+R^0
\ee
(see the definition of $g_+$ in the list (A.5)).

Thus we have

\ses\ses

PROPOSITION.
{\it
The velocity of the point of intersection of the zero-phase surface with the $R^1$--axis
is isotropic in accordance with }(C.6).
{\it The vertex
 moves along the  axis with the constant velocity
\be
 v= g_+.
\ee
}

\ses
\ses

It will be noted that  under the particular conditions (C.5)
the  vanishing $F=0$ for the
$\cE^{SR}_g$--space  implies the precise
isotropic value (C.6) (see (A.12)).


Let us put
\be
R_{\perp}=\sqrt{(R^2)^2+(R^3)^2}
\ee
obtaining
\be
\mR=\sqrt{(R^1)^2+(R_{\perp})^2}.
\ee
We find
\be
R^0={\rm const}:\qquad \D{R^1}{R_{\perp}}=
-\fr g2\fr{R_{\perp}}{h\mR+\fr g2R^1},
\ee
so that
\be
R^0={\rm const}:\qquad \D{R^1}{R_{\perp}}\biggl|_{R_{\perp}=0}=0
\ee
and
\be
R^0={\rm const}:\qquad \DD{R^1}{R_{\perp}}\biggl|_{R_{\perp}=0}=-\fr g2g_+\fr 1{R^1}.
\ee
The equation (11.4) can be resolved for $R^1$:
\be
R^1=hR^0-\fr g2\sqrt{(R^0)^2+(R_{\perp})^2}.
\ee
The
  {\it asymptotic cone} $\cal C$ is determined by the equation
\be
R^1=-\fr g2R_{\perp}
\ee
and simultaneously presents the front at $R^0=0$. The asymptotic cone $\cal C$
is the same for any case of  zero-phase wave.
When $R^0\to\iy$,
the right-hand part in (C.12) tends to zero
and, therefore,
{\it the front  is flattening around the vertex.}

In the approximation $O_1(g)$, we put $h=1$ and reduce the formulae (C.2) and (C.4) to merely
\be
(R^0-\fr12g\mR)k_0-k_aR^a =0
\ee
and
\be
R^0=R^1+\fr12g\mR.
\ee
The equality (C.6) takes on the form
\be
R^1=(1-\fr12g)R^0,
\ee
and instead of (C.13) we get
\be
R^1=R^0-\fr g2\sqrt{(R^0)^2+(R_{\perp})^2}.
\ee


Below,
the zero-phase wave fronts determined by the equation (C.1)--(C.6)
are shown at various circumstances.

\pgbrk


\begin{figure}
\footnotesize
\psfrag{R[1]}{\rm$R^1$}
\psfrag{R[2]}{\rm$R^2$}
\psfrag{R[3]}{\rm$R^3$}
\centering
\vbox to \textheight { \vfil
\includegraphics[width=10cm]{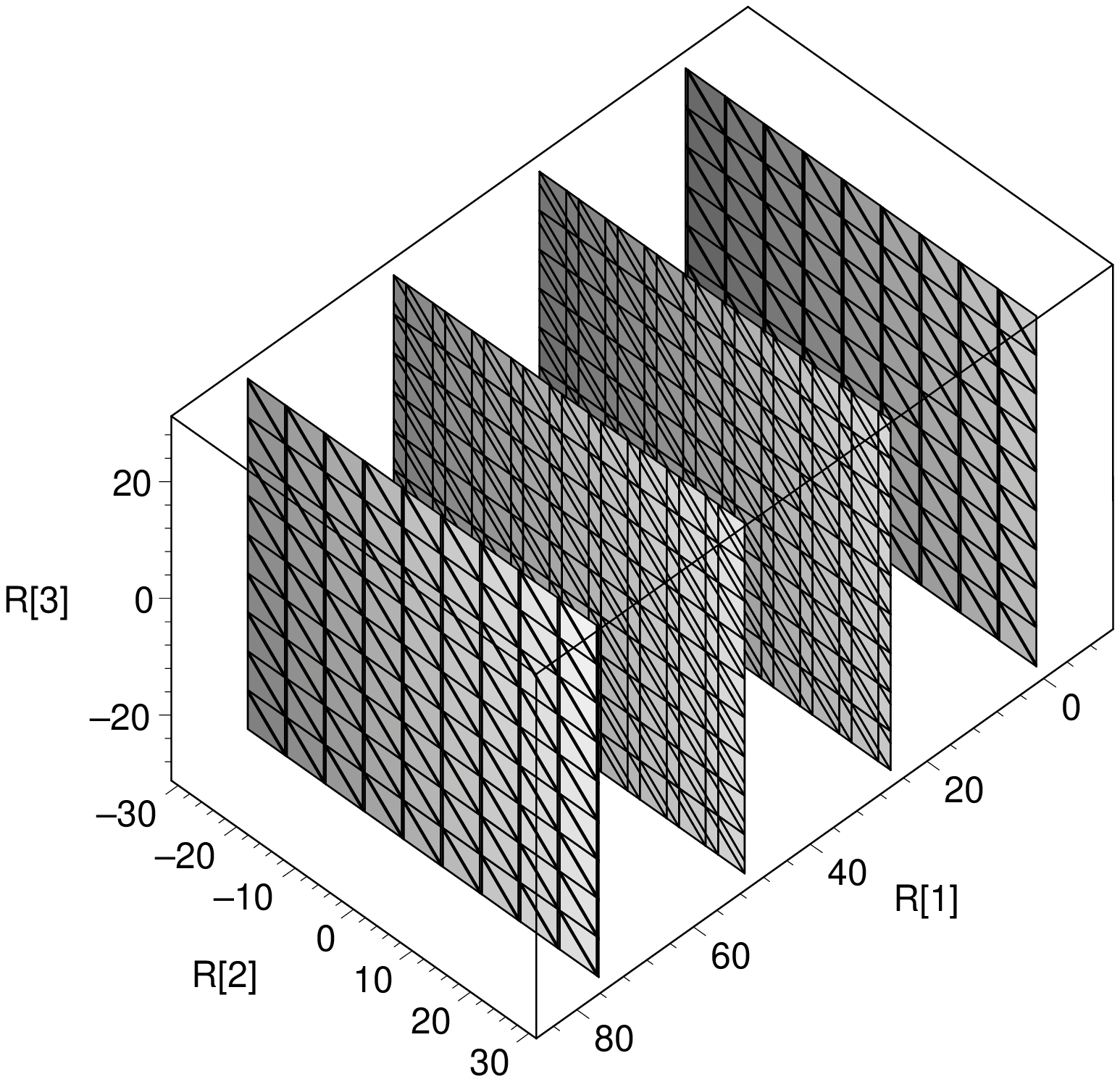}
\vfil
\vspace{0.2cm}
\includegraphics[width=10cm]{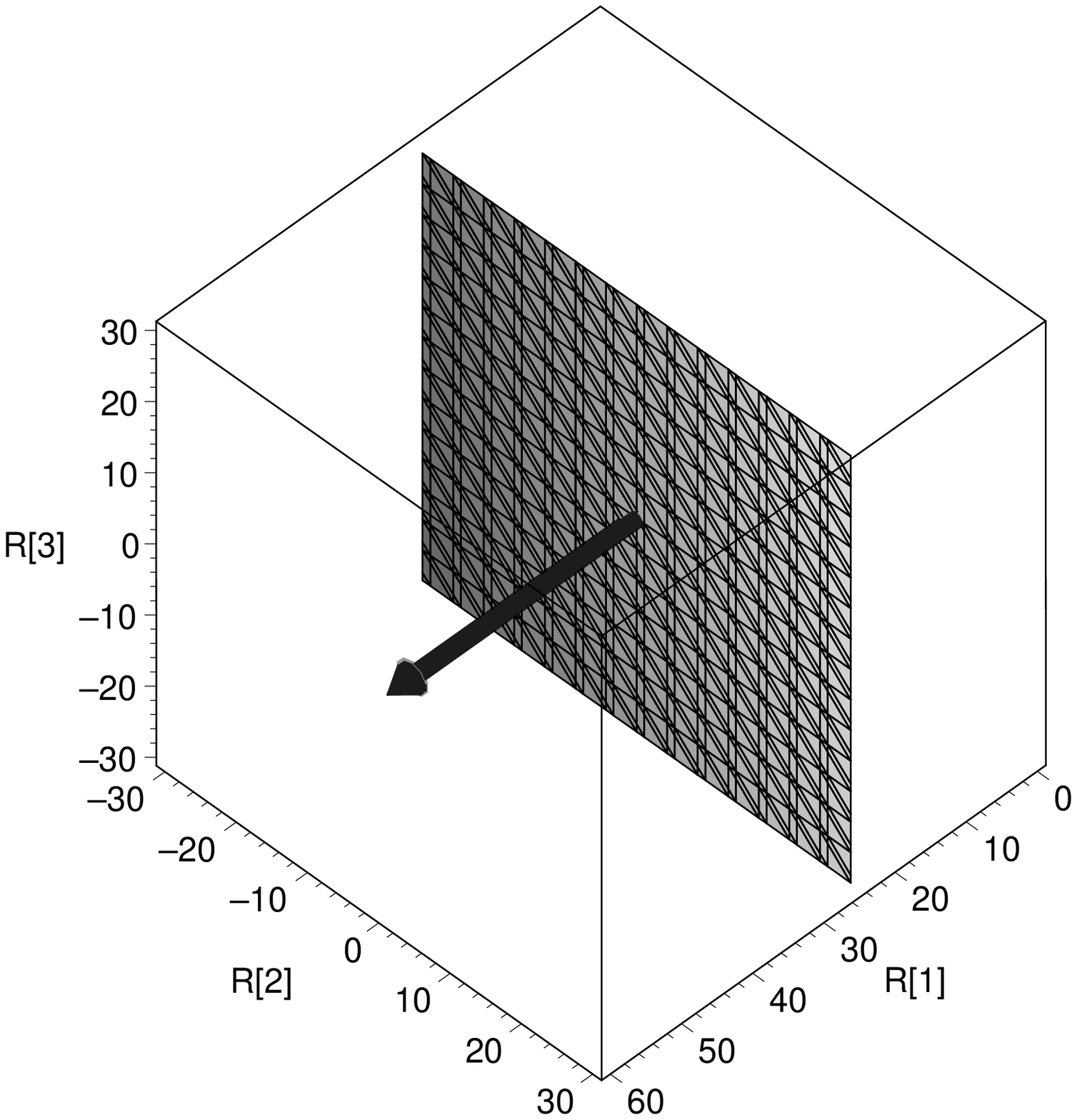}
\vfil }
\caption{\small The fronts are
 pictured at the conventional pseudoeuclidean case, which is the trivial case $g=0$
of the Finslerian approach under study. In this case the fronts are plane}

\end{figure}


\begin{figure}
\footnotesize
\psfrag{R[1]}{\rm$R^1$}
\psfrag{R[2]}{\rm$R^2$}
\psfrag{R[3]}{\rm$R^3$}
\centering
\vbox to \textheight  { \vfil
\includegraphics[width=10cm]{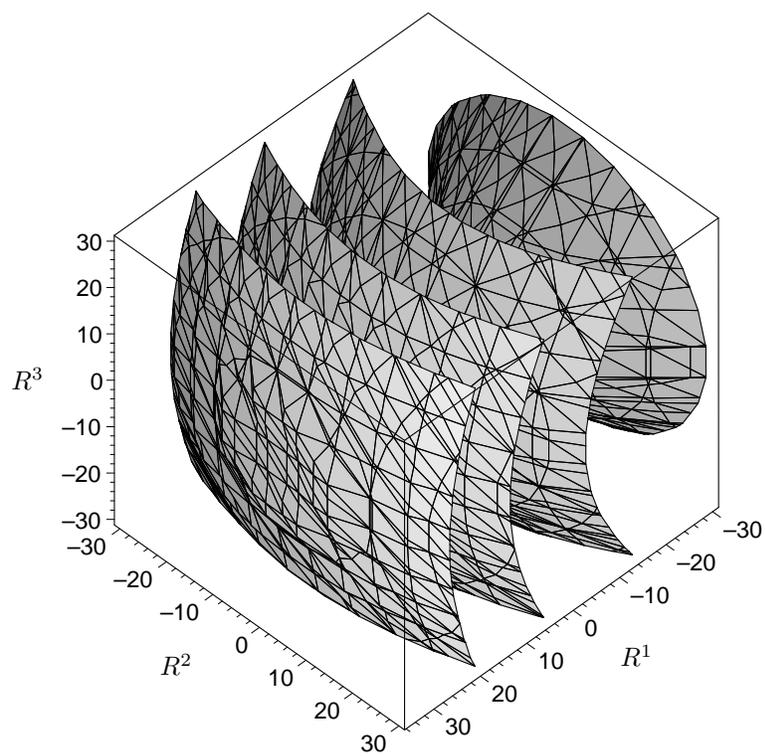}
\vfil
\vspace{0.2cm}
\includegraphics[width=10cm]{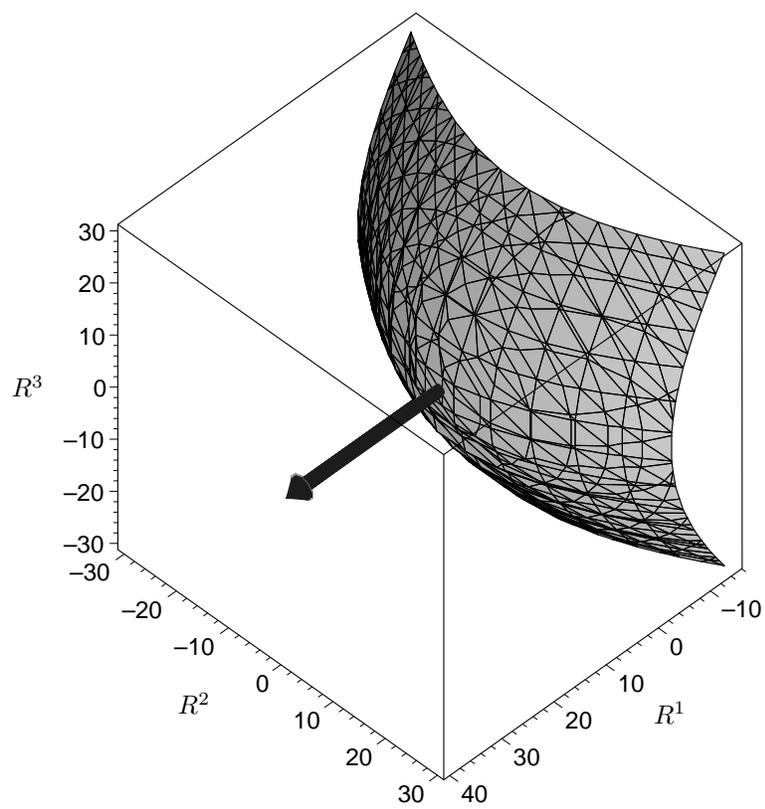}
\vfil }
\caption{\small The case
$g=2$}

\end{figure}


\clearpage

\begin{figure}
\footnotesize
\psfrag{R[1]}{\rm$R^1$}
\psfrag{R[2]}{\rm$R^2$}
\psfrag{R[3]}{\rm$R^3$}
\centering
\vbox to \textheight  { \vfil
\hbox to \textwidth
{
\includegraphics[width=7cm]{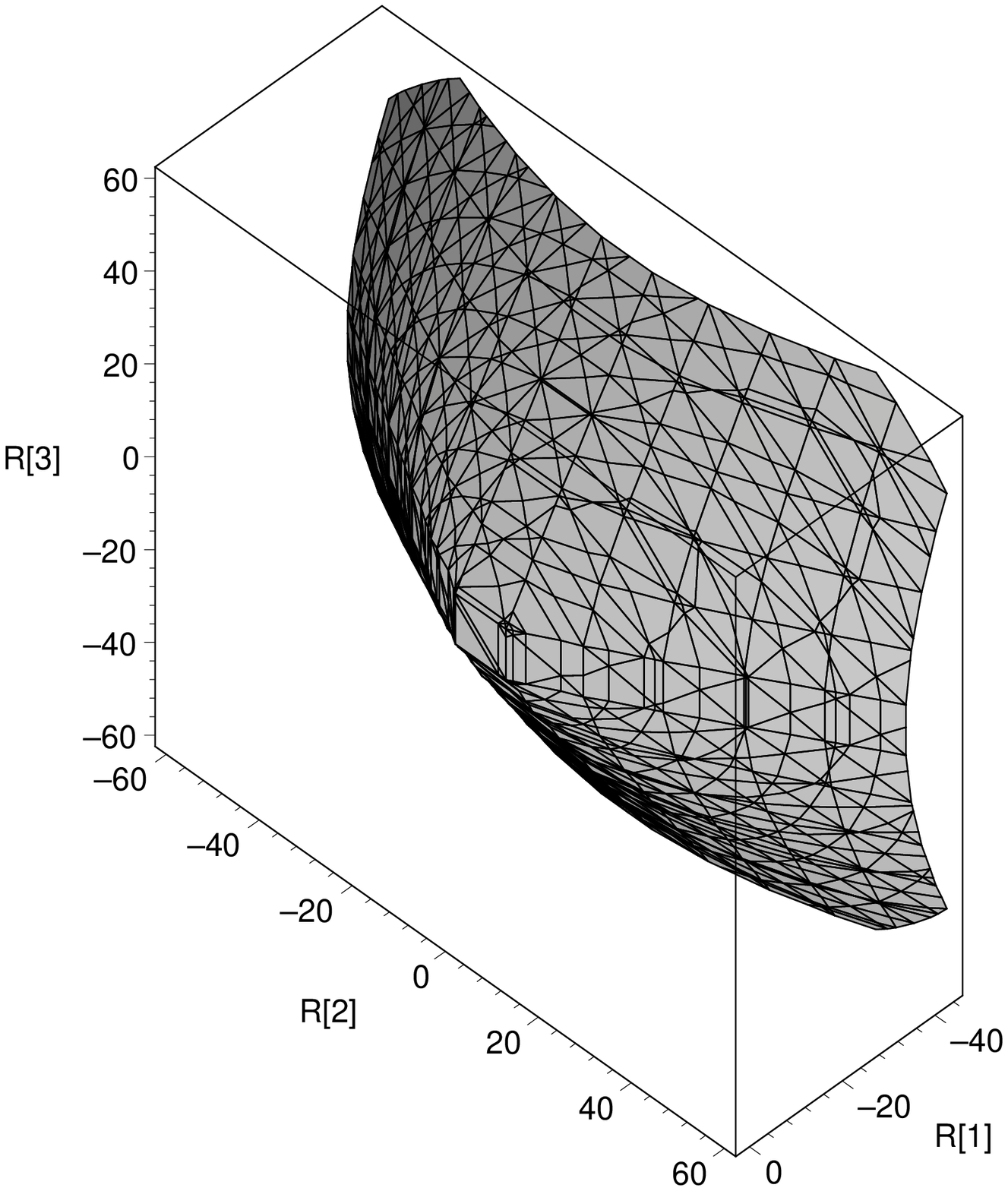}
\includegraphics[width=7cm]{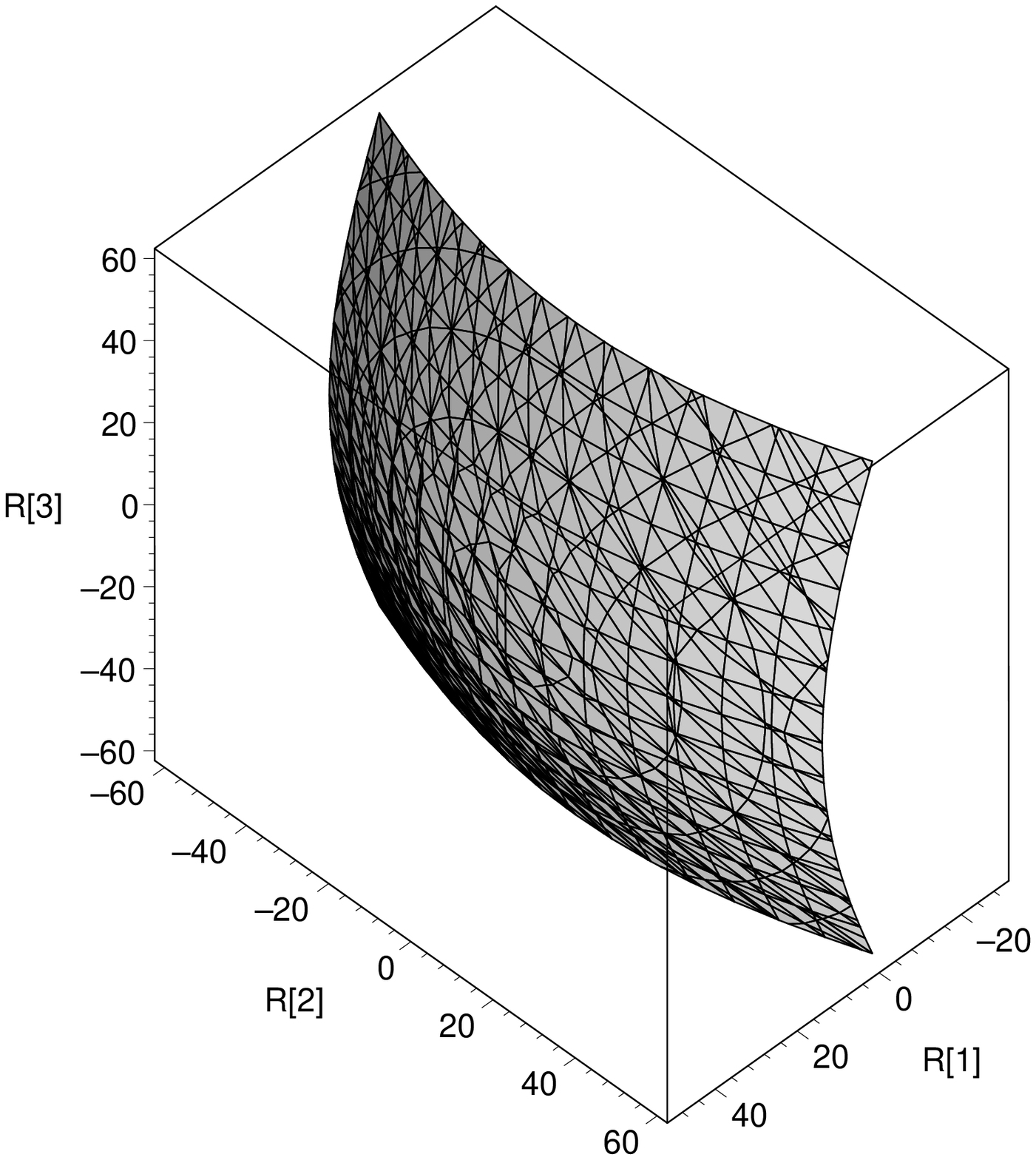}
}
\vfil
\hbox to \textwidth
{
\includegraphics[width=7cm]{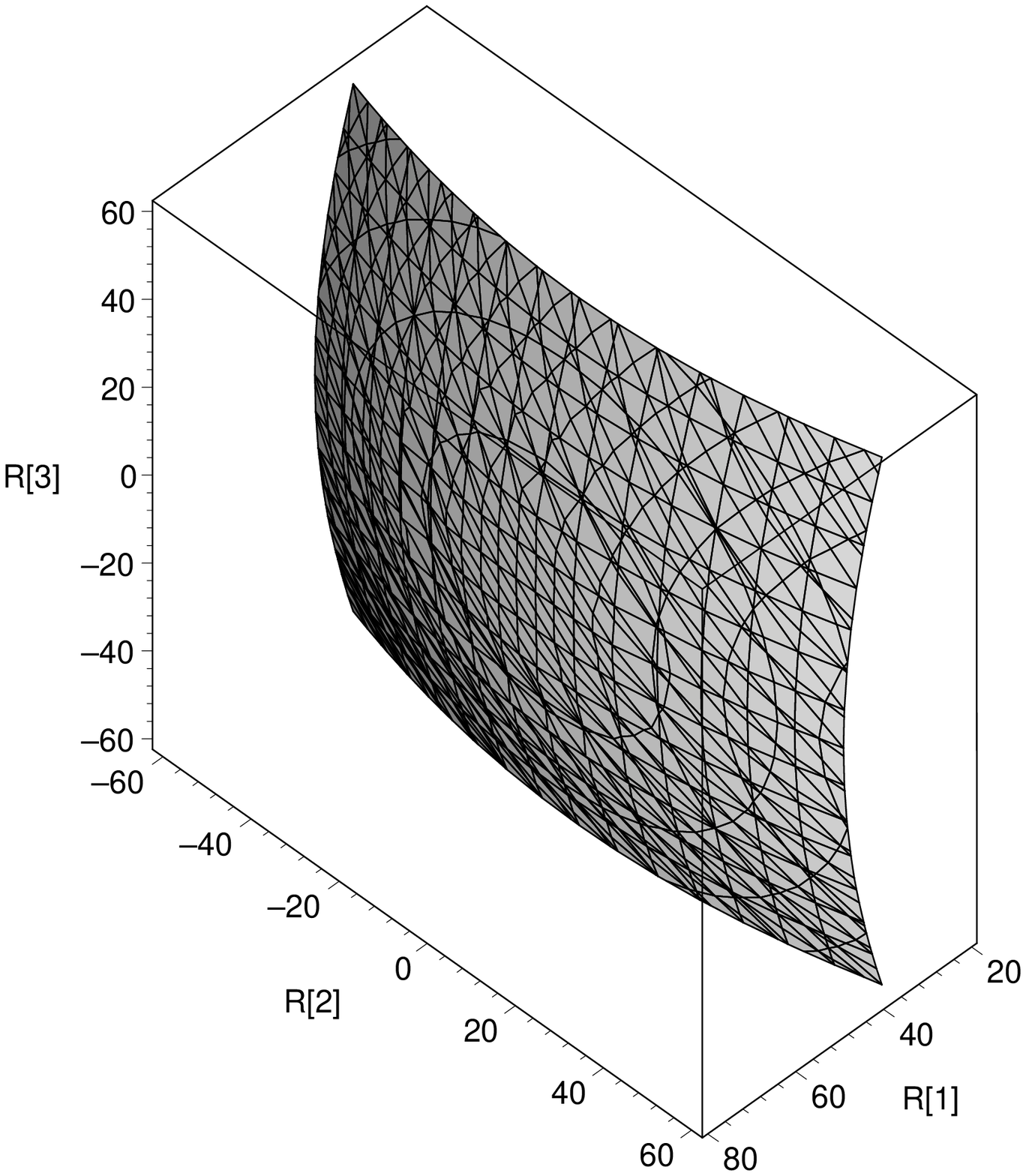}
\includegraphics[width=7cm]{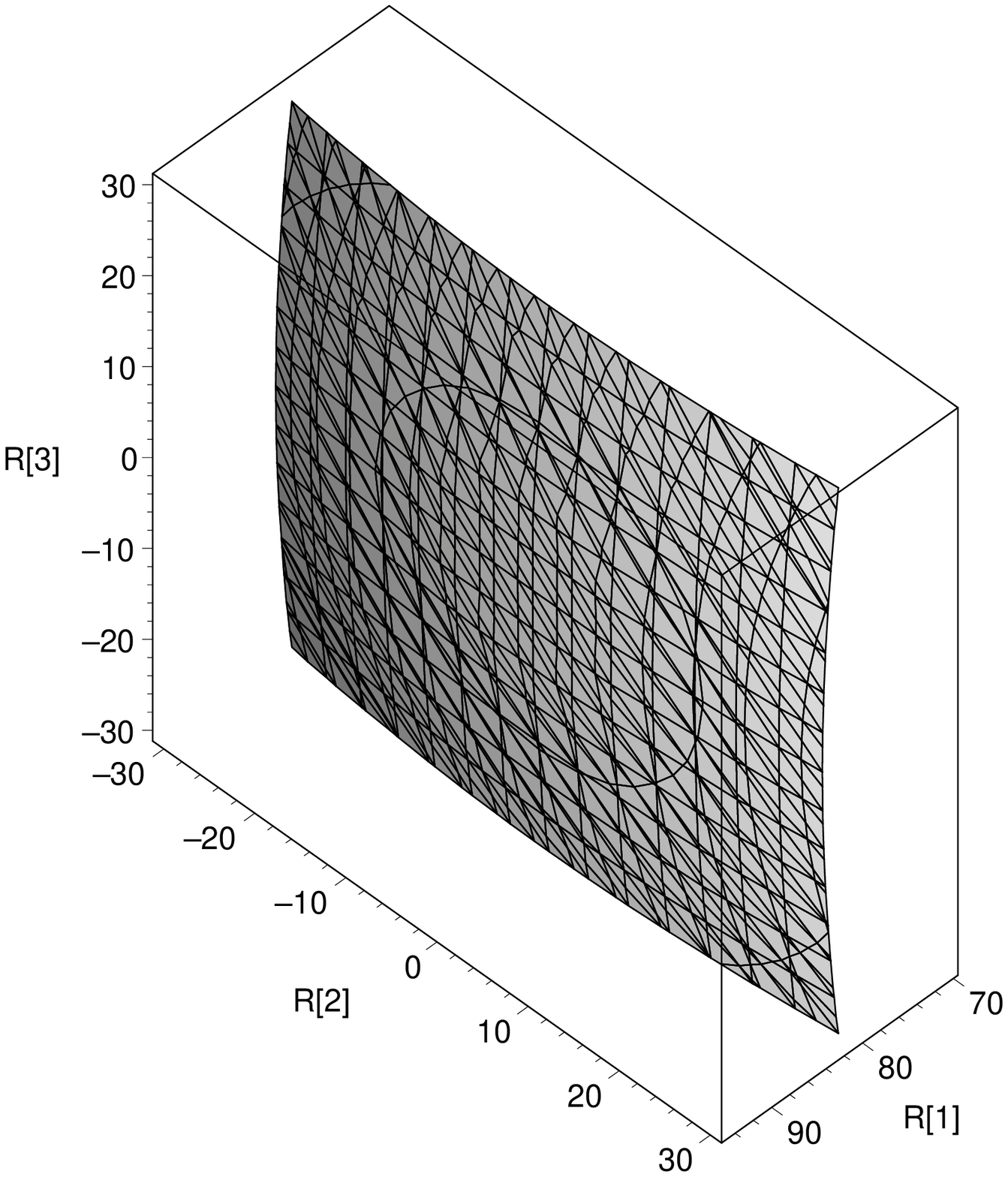}
}
\vfil }
\caption{The fronts are shown for the   value
 $g=1.2$. The higher case pictures correspond to the time momenta
 $R^0=0$ and $R^0=50$, whereas the lower case pictures reflect the cases
 $R^0=100$ and $R^0=150$}
\end{figure}

\begin{figure}
\footnotesize
\psfrag{R[1]}{\rm$R^1$}
\psfrag{R[2]}{\rm$R^2$}
\psfrag{R[3]}{\rm$R^3$}
\centering
\vbox to \textheight
{ \vfil \hbox to \textwidth {
\includegraphics[width=7cm]{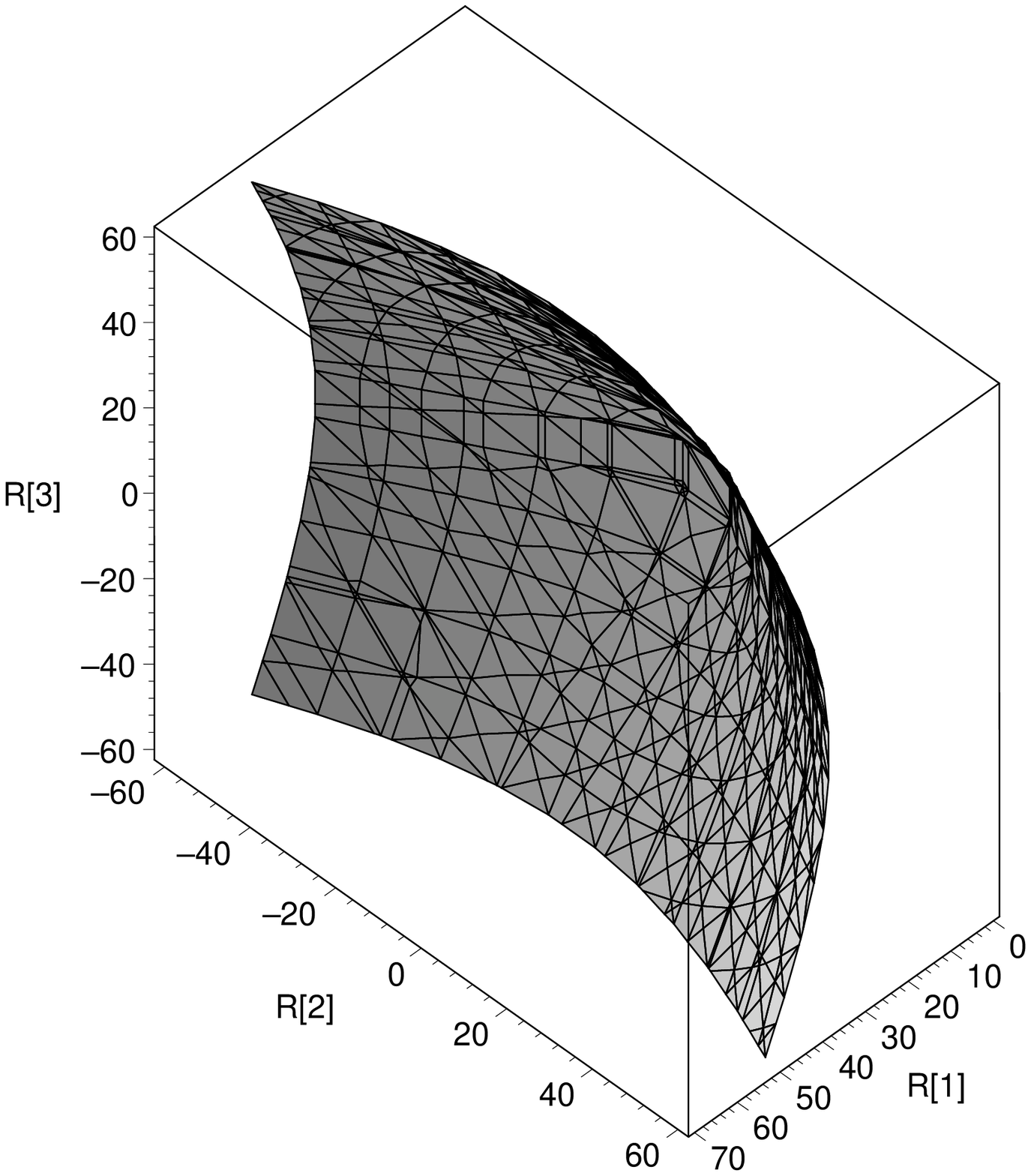}
\includegraphics[width=7cm]{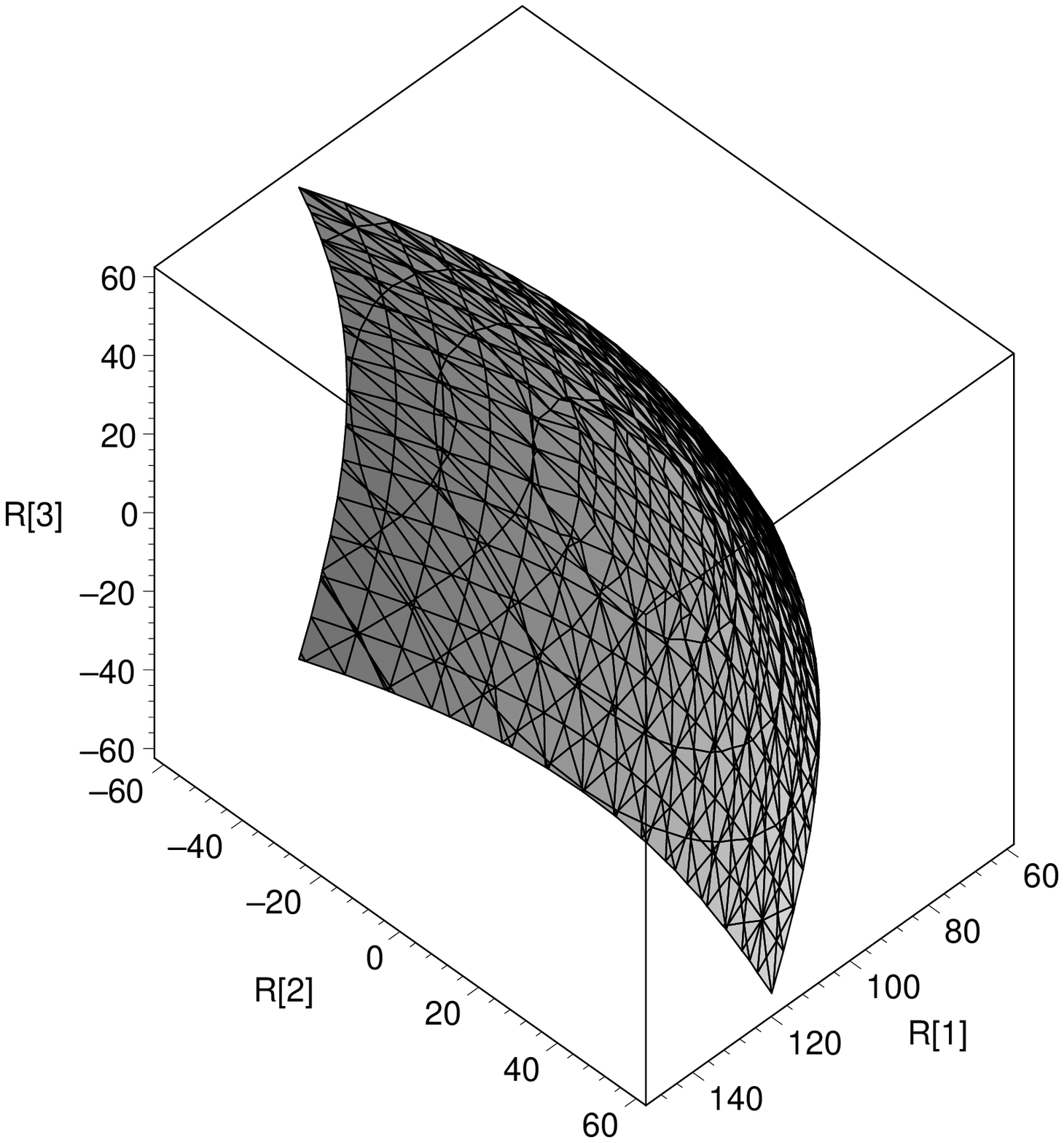}
} \vfil \hbox to \textwidth {
\includegraphics[width=7cm]{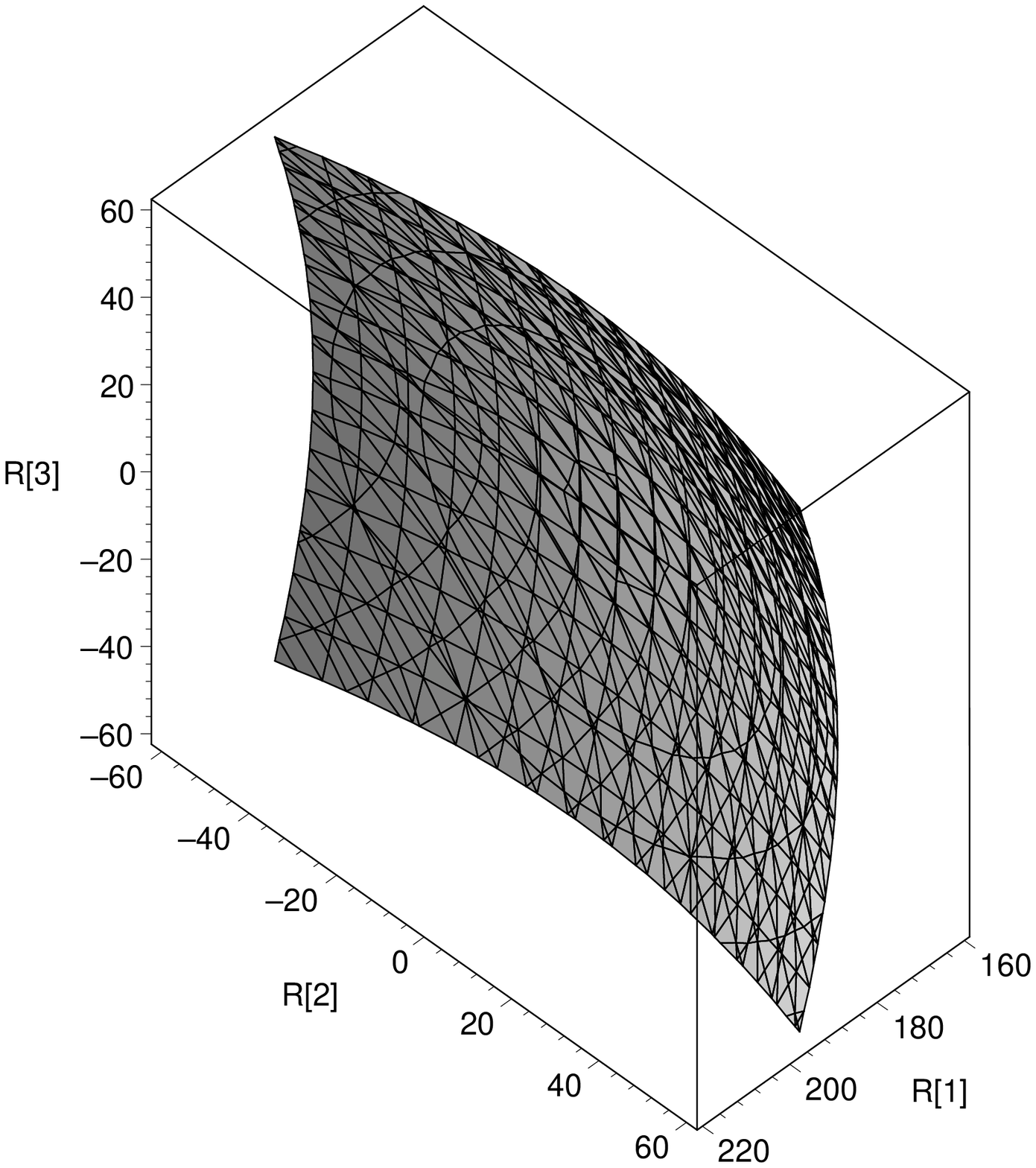}
\includegraphics[width=7cm]{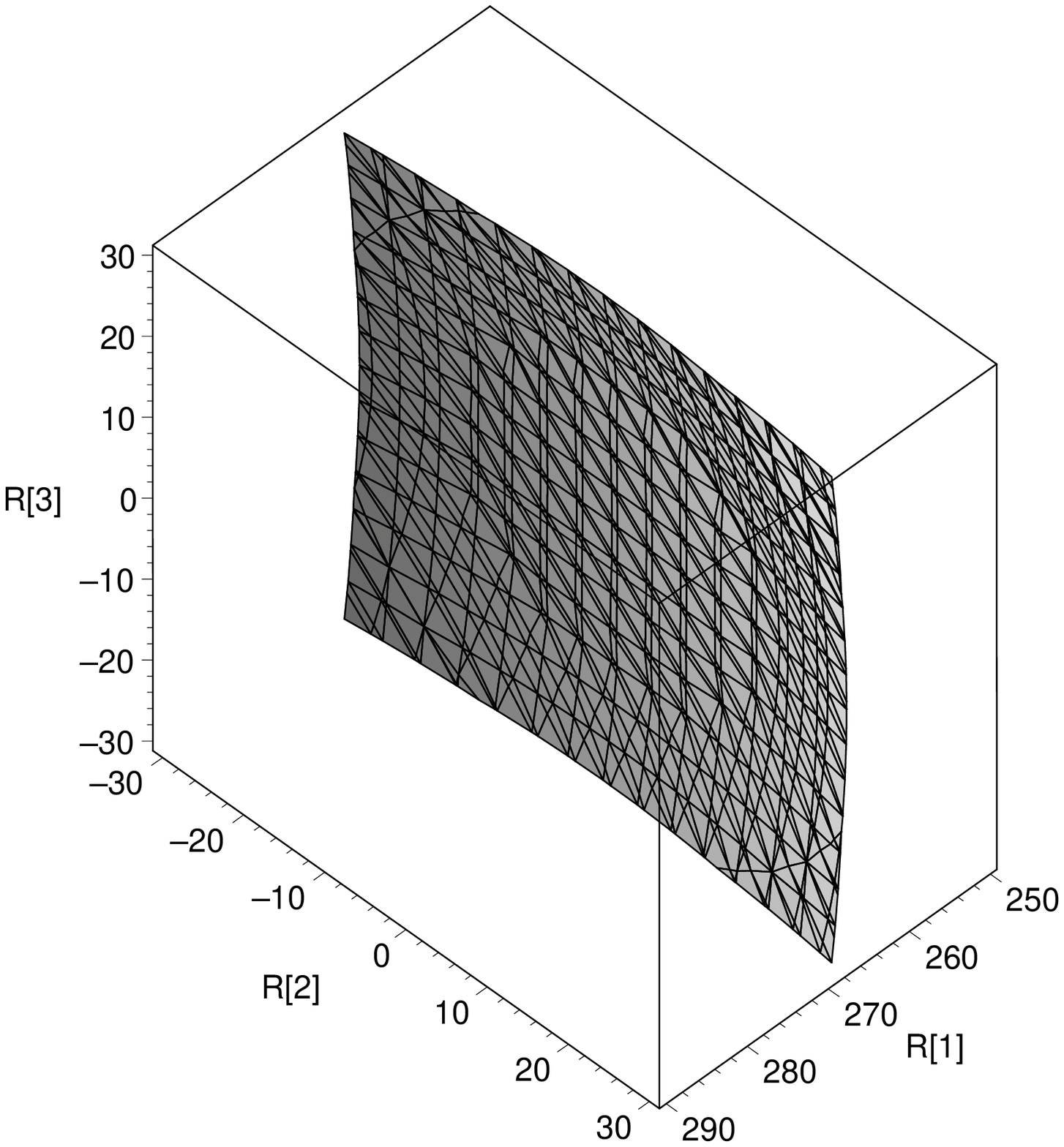}
} \vfil }
\caption{The fronts are shown for the   value
 $g=-1.2$. The higher case pictures correspond to the time momenta
 $R^0=0$ and $R^0=50$, whereas the lower case pictures reflect the cases
 $R^0=100$ and $R^0=150$}
\end{figure}

\clearpage




\begin{figure}
\footnotesize
\psfrag{R[1]}{\rm$R^1$}
\psfrag{R[2]}{\rm$R^2$}
\psfrag{R[3]}{\rm$R^3$}
\centering
\vbox to \textheight  {
\vfil
\includegraphics[width=10cm]{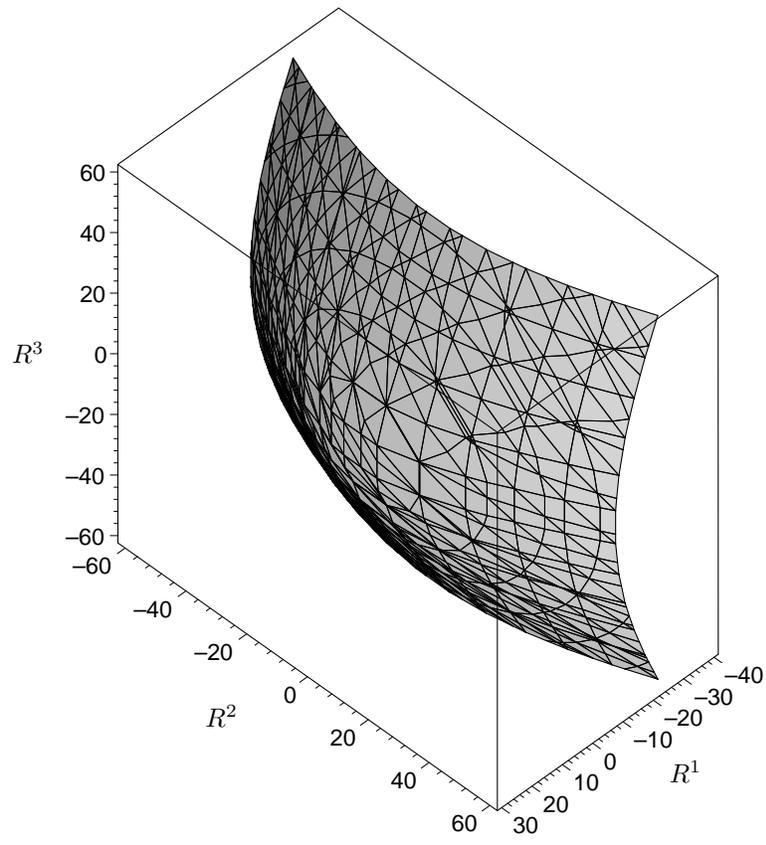}
\vfil
\vspace{0.2cm}
\includegraphics[width=10cm]{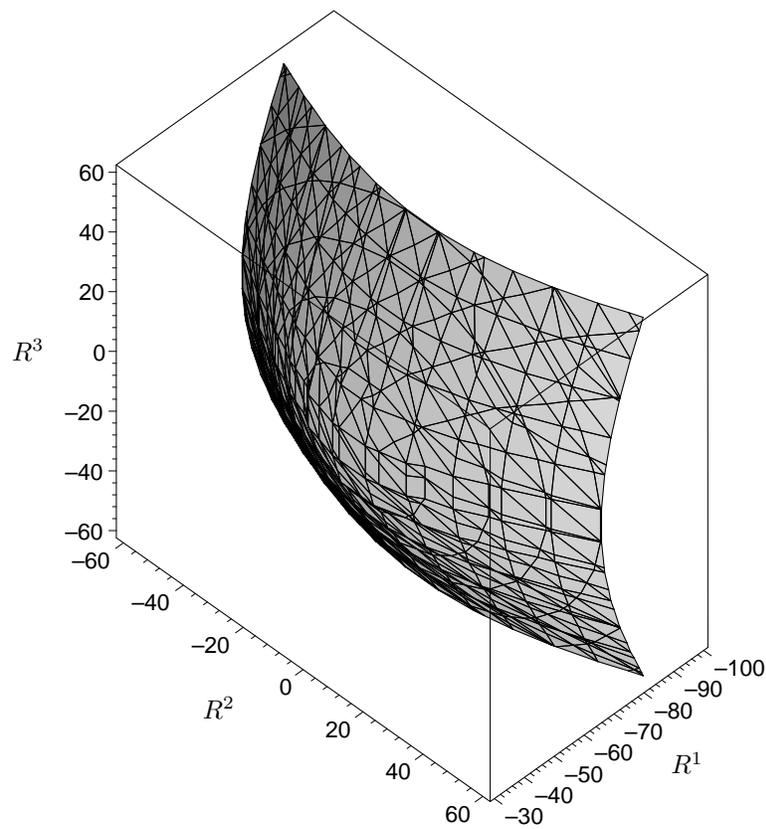}
\vfil }
\caption{\small The case
 $g=0.8$;
 $R^0=25$ at the top picture and $R^0=-25$ at the bottom picture }
\end{figure}

\clearpage

\begin{figure}
\psfrag{R1}{\raisebox{-0.2cm}{\rm$R^1$}}
\psfrag{R2}{\rm$R^2$}
\centering
\vbox to \textheight {\vfil
\includegraphics[width=8cm]{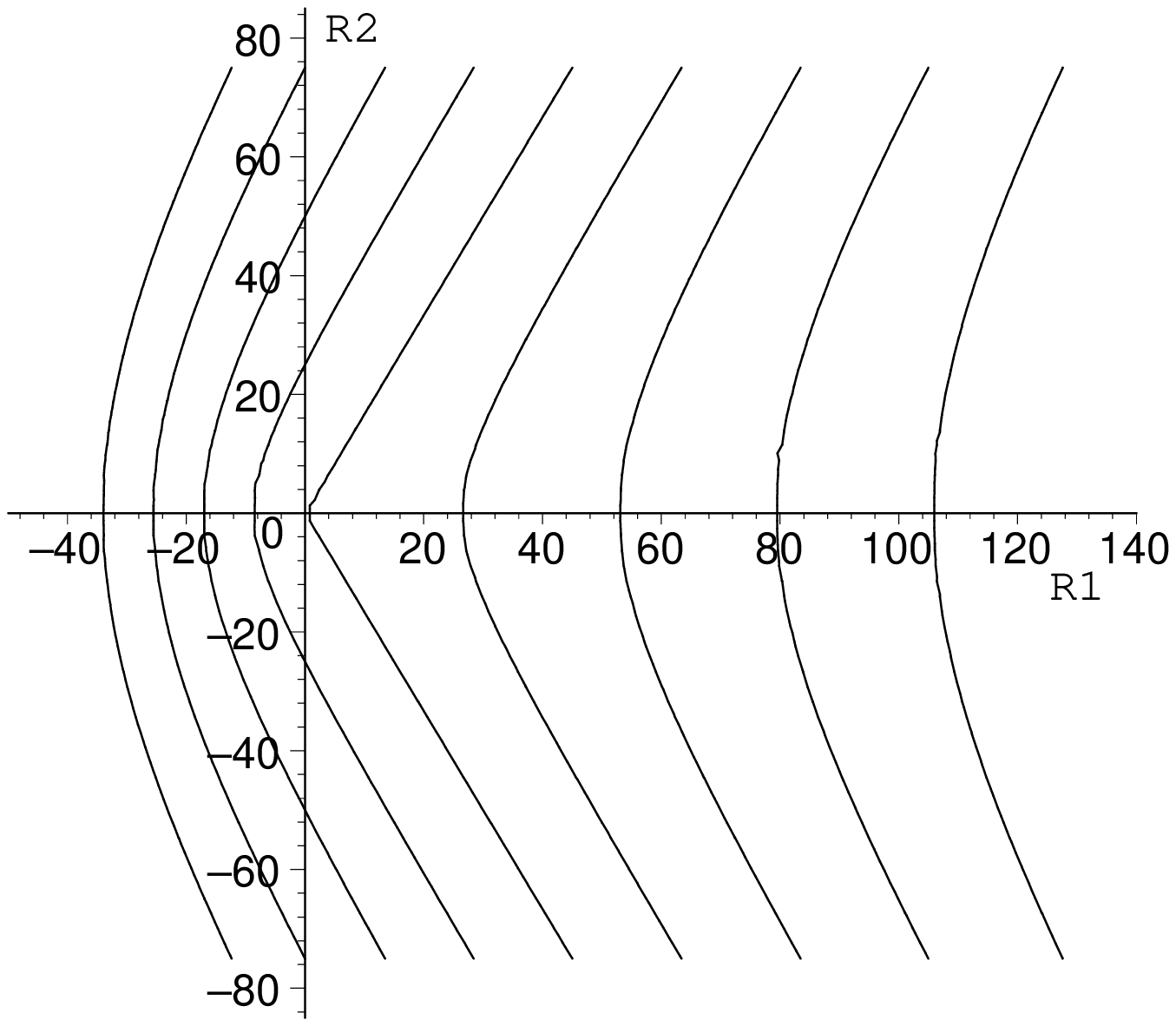}
\vfil
\includegraphics[width=8cm]{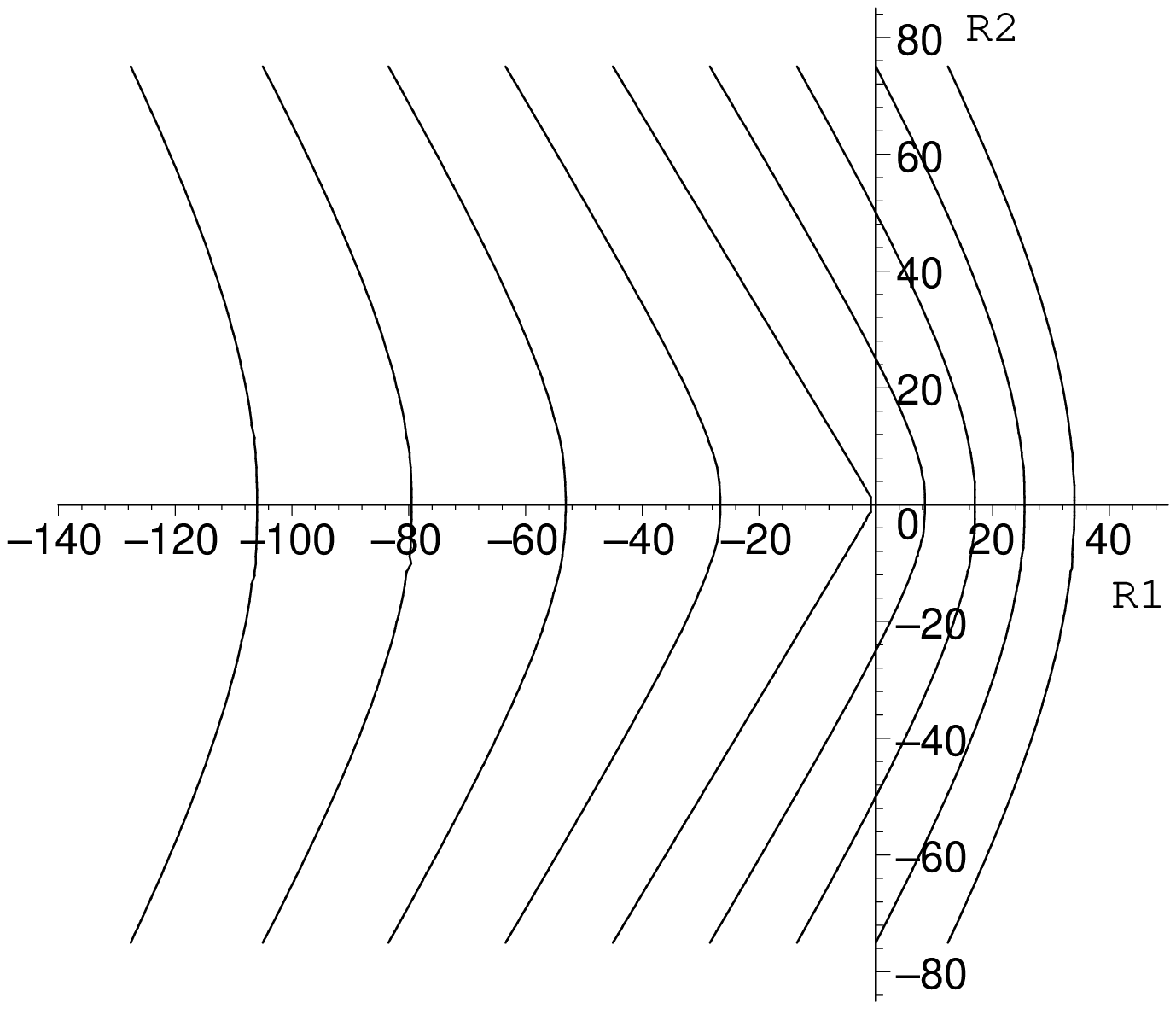}
\vfil }
\caption{\small With $R^0=-60,-45,-30,-15,0,15,30,45,60$, the fronts are simulated
in the
two-dimensional case, when $R^3=0$, of the considered zero-phase
wave;
$g=-1.2$ at the top picture and $g=1.2$ at the bottom picture }
\end{figure}

\clearpage

We preluded with Fig. 1 which symbolized the ordinary pseudoeuclidean plane wave,
that is, the $(g=0)$--case.

In Fig. 2 the fronts
were depicted at various values of the parameter
$g\ne 0$, showing that
the fronts  are non-plane.
In the higher figures the values
$R^0=0;15;30;45$
were  subsequently be taken.
Lower figures symbolized the front at the initial momentum
$R^0=25$
and the arrow shows the direction of propagation.
At
$R^0=0$
the vertex is a conic point.
We observe how the front is flattening in a vicinity of vertex as soon as the value
of $R^0$ is increasing.

In Figs. 3 and 4 the fronts were shown separately at the values $R^0=0,50,100,150$
for positive and definite values of $g$.
Additionally,
Fig. 5 shows two kindred fronts at two opposite values of the temporal coordinate
$R^0$.
Finally, Fig. 6 showed the succession of the fronts in the $2d$-case where the
vanishing $R^3=0$ has been assumed.

When $g<0$ the fronts move in the direction pointed by the vertex; the opposite direction
of motion takes place at $g>0$. In each figure the fronts move in the positive direction of the
$R^1$--axis.

\ses\ses

\setcounter{sctn}{4} \setcounter{equation}{0}

{\nin\bf Appendix D. Conformal flatness of the space $\cE_g^{SR}$}

\ses \ses

Let us use the scalar
\be
\Ka(g;R)=\lf(\fr12F^2(g;R)\rg)^{h-1}
\equiv\bigl(\ka(g;R)\bigr)^2
\ee
and introduce the new tensor  $s_{pq}(g;R)$ according to
\be
g_{pq}(g;R)=\fr1{\Ka(g;R)}s_{pq}(g;R).
\ee
Taking the associated Christoffel symbols
\be
S_p{}^t{}_r=
\fr12s^{ts}\lf(\D{s_{ps}}{R^r}+\D{s_{rs}}{R^p}-\D{s_{pr}}{R^s}\rg),
\ee
we can straightforwardly evaluate the curvature tensor
\be
M_t{}^u{}_{rs}\eqdef\D{S_t{}^u{}_r}{R^s}-\D{S_t{}^u{}_s}{R^r}+
S_t{}^w{}_rS_w{}^u{}_s-S_t{}^w{}_sS_w{}^u{}_r.
\ee
The result proves to be zero:
\be
M_t{}^u{}_{rs}\equiv0.
\ee
Thus we get

\ses\ses

{PROPOSITION}. {\it The space
$\cE_g^{SR}$
is conformally flat. The conformal multiplier is a function of the FMF according
to} (D.1)--(D.2).

\ses\ses

Therefore, the transformation
\be
r^i= \rho^i(g;R)
\ee
should exist such that in terms of the coefficients
\be
\rho^i_q(g;R) \eqdef\D{\rho^i(g;R) }{R^q}
\ee
the identity
\be
\fr1{h^2}s_{pq}(g;R)=\rho^i_p(g;R)\rho^j_q(g;R)e_{ij}
\ee
holds fine
($e_{ij}={\rm diag}(1,-1,-1,-1)$ is a  pseudoeuclidean metric tensor).
A careful consideration leads to

\ses\ses

{PROPOSITION}. {\it The conformal transformation} (D.6)
{\it is given by the components
}
\be
\rho^0(g;R) =\fr1h\ka(g;R)j(g;R)\lf(R^0-\fr12g\mR\rg),
\quad \rho^a(g;R) =\ka(g;R)j(g;R)R^a.
\ee

\ses
\ses

The identity
\be
\biggl(R^0-\fr12g\mR\biggr)^2-h^2\mR^2=-B(g;R)
\ee
is useful to apply when verifying this Proposition.


The transformation inverted to (D.6) is given by the law
\be
R^p=\eta^p(g;r)
\ee
with the functions
\be
\eta^0(g;r)=\nu(g;r)\lf(hr^0+\fr12g\mr \rg)/j(g;r),
\quad\eta^a(g;r)=\nu(g;r)r^a/j(g;r),
\ee
where
\be
\nu(g;r)=\lf(\fr{h}{\sqrt2}S(g;r)\rg)^{(h-1)/h}=\fr1{\ka(g;R)}
\ee
\ses
and
\be
j(g;r)=\lf(\fr{r^0-\mr}{r^0+\mr}\rg)^{-G/4}.
\ee

Using the derivatives
\be
\eta_i^p(g;r)\eqdef\D{\eta^p(g;r)}{r^i}              ,
\ee
we may rewrite (D.8)  in the backward direction as follows:
\be
\eta_i^p(g;r)\eta_j^q(g;r)s_{pq}(g;R)=e_{ij}.
\ee

It is convenient to use the notation
\be
\ga=h-1.
\ee


{\it In the time-like case} we get
\be
\rho^0(g;R) =\fr1{h}\ka(g;R)j(g;w)E(g;w)R^0,
\quad \rho^a(g;R) =\ka(g;R)j(g;w)R^a,
\quad R\in S_g^+,
\ee
where
\be
E(g;w)=1-\fr12gw.
\ee
The function $j$ entering (D.37) has been defined by  (A.13).
We obtain
the identity
\be
\biggl(E(g;w)\biggr)^2-h^2w^2=Q(g;w)
\ee
and the representations
\be
\rho^0_0(g;R) =\lf(\ga E(1-gw)+Q-\fr12gwE\rg)\fr{j\ka}{Qh^2},
\ee
\bigskip
\be
\rho^0_a(g;R) =\lf(-\ga E+\fr{g(E-Q)}{2w}\rg)\fr{j\ka w_a}{Qh^2},
\ee
\bigskip
\be
\rho^a_0(g;R) =\lf(\ga(1-gw)-\fr12gw\rg)\fr{j\ka w^a}{Qh},
\ee
\bigskip
\be
\rho^a_b(g;R) =\lf(-\ga w_bw^a+Q\delta^a_b+\fr{gw_bw^a}{2w}\rg)\fr{j\ka}
{Qh},
\ee
\ses\\
from which the identity
\bigskip\\
\be
\de_{cd}\rho^c_a\rho^d_b=\Biggl(\de_{ab}+\fr{w_aw_b}{Qw}\Bigl(-\ga w+\fr12g\Bigr)
\Bigl(2-\fr{\ga w^2}Q+\fr{gw}{2Q}\Bigr)\Biggr)\fr{(j\ka)^2}{h^2}
\ee
follows.
Some simplification is applicable to find
\be
\rho^0_0(g;R)=\lf(E(1-gw)-h\,w^2\rg)\fr{j\ka}{Qh},
\ee
\bigskip
\be
\rho^0_a(g;R)=\lf(h-E\rg)\fr{j\ka w_a}{Qh} ,
\ee
\bigskip
\be
\rho^a_0(g;R)=\lf(hQ-E+h\,w^2\rg)\fr{j\ka w^a}{Qh},
\ee
\bigskip
\be
\rho^a_b(g;R)=\Biggr(Q\delta^a_b+\lf((1-h)w+\fr g2\rg)
\fr{w_bw^a}w\Biggr)\fr{j\ka}{Qh},
\ee
together with
\be
\de_{cd}\rho^c_a(g;R)\rho^d_b(g;R)=\Biggl[\de_{ab}+
\lf(g+(1-h)w+(h^2-h)\fr{w(1+w^2)}Q\rg)
\fr{w_aw_b}{Qw}\Biggr](j\ka)^2.
\ee


Also,
\be
{\rm det}(\rho^i_p)=(j\ka)^N,
\qquad {\rm det}(\rho^a_b)=\lf(E-h\,w^2\rg)\fr{(j\ka)^{N-1}}Q,
\ee
\bigskip
\be
w^b\rho^a_b=\lf(E-h\,w^2\rg)\fr{j\ka w^a}Q,
\qquad w_a\rho^a_b=\lf(E-h\,w^2\rg)\fr{j\ka w_b}Q,
\ee
\bigskip
\be
\partial_q\ka = \fr{\ga\ka}{F^2}R_q
\ee
$(N=4)$.

For the reciprocal coefficients defined in accordance with (D.15) we find the representations
\ses\\
\be
\eta^0_0(g;r) =\lf(E-h\,w^2\rg)\fr{h}{j\ka Q},
\ee
\bigskip
\be
\eta^0_a(g;r) =\lf(E-h\rg)\fr{w_ah}{j\ka Q},
\ee
\bigskip
\be
\eta^a_0(g;r) =\lf(E-hw^2-hQ\rg)\fr{w^ah}{j\ka Q},
\ee
\bigskip
\be
\eta^a_b(g;r) =\Biggr(Q\delta^a_b+\lf((E-h)w-\fr12gQ\rg)
\fr{w_bw^a}w\Biggr)\fr{h}{j\ka Q},
\ee
\ses\\
and the identities
\be
R_p\eta^p_0=\fr{EF^2h}{R^0j\ka Q},
\qquad R_p\eta^p_a=-\fr{h^2F^2w_a}{R^0j\ka Q},
\ee
\bigskip
\be
\D{\lf(\ka^{-3}\sqrt{|\det(s_{rs})|}\,\eta^p_0\rg)}{R^p}=
-6\ga\fr{EF^2j^3h}{R^0Q}.
\ee
\bigskip


{\it In the space-like region},
\be
\rho^0(g;R) =\fr1{h}\ka(g;R)j(g;k)f(g;k)q,
\quad \rho^a(g;R) =\ka(g;R)j(g;k)R^a,
\qquad R\in\cR^+_g,
\ee
where
\be
f(g;k)=k-\fr12g,
\ee
so that
\be
h^2-\biggl(f(g;k)\biggr)^2=L(g;k)
\ee
\ses\\
and
\be
\rho^0_0(g;R)=\Bigl(-\ga(k-g)f+L+\fr12gf\Bigr)\fr{j\ka}{Lh},
\ee
\bigskip
\be
\rho^0_a(g;R)=\Bigl(\ga f-\fr12g(L+kf)\Bigr)\fr{j\ka n_a}{Lh},
\ee
\bigskip
\be
\rho^a_0(g;R)=\Bigl(-\ga(k-g)+\fr12g\Bigr)\fr{j\ka n^a}L,
\ee
\bigskip
\bigskip
\be
\rho^a_b(g;R)=\Bigl(L\delta^a_b+(\ga-\fr12gk)n^an_b\Bigr)\fr{j\ka}L
\ee
\ses
or
\be
\rho^0_0(g;R)=\Bigl(h-(k-g)f\Bigr)\fr{j\ka}L,
\ee
\bigskip
\be
\rho^0_a(g;R)=\Bigl(f-kh\Bigr)\fr{j\ka n_a}L,
\ee
\bigskip
\be
\rho^a_0(g;R)=\Bigl(f-(k-g)h\Bigr)\fr{j\ka n^a}L,
\ee
\bigskip
\bigskip
\be
\rho^a_b(g;R)=\Bigl(L\delta^a_b+(h-1-\fr12gk)n^an_b\Bigr)\fr{j\ka}L,
\ee
\bigskip
\bigskip
\be
\det(\rho^i_p)=(j\ka)^N,
\qquad \det(\rho^a_b)=\lf(h-fk\rg)\fr{(j\ka)^{N-1}}L
\ee
$(N=4)$.
\ses\\
Also,
\be
n^b\rho^a_b=\lf(h-fk\rg)\fr{j\ka n^a}L,
\qquad w_a\rho^a_b=\lf(h-fk\rg)\fr{j\ka n_b}L,
\ee
\bigskip
\be
\partial_q\ka = -\fr{\ga\ka}{F^2}R_q.
\ee



The reciprocal coefficients
are obtainable to read
\be
\eta^0_0(g;r) =\lf(h-fk\rg)\fr1{j\ka L},
\ee
\bigskip
\be
\eta^0_a(g;r) =\lf(kh-f\rg)\fr{n_a}{j\ka L},
\ee
\bigskip
\be
\eta^a_0(g;r) =\lf((k-g)h-f\rg)\fr{n^a}{j\ka L},
\ee
\bigskip
\be
\eta^a_b(g;r) =\Biggl(L\delta^a_b+
\lf(h-(k-g)f-L\rg)n_bn^a\Biggr)\fr1{j\ka L},
\ee
entailing
\be
R_p\eta^p_0=-\fr{fF^2}{qj\ka L},
\qquad R_p\eta^p_a=\fr{hF^2n_a}{qj\ka L},
\ee
\bigskip
\be
\D{\lf(\ka^{-3}\sqrt{|\det(s_{rs})|}\,\eta^p_0\rg)}{R^p}=
6\ga\fr{fF^2j^3}{qL}.
\ee


\ses\ses

\setcounter{sctn}{5} \setcounter{equation}{0}

{\nin\bf Appendix E. $O_1(g)$--approximations}

\ses \ses


Considering the parameter $g$ to be small:
\be
|g|\ll 1
\ee
and performing expansions with respect to the parameter,
we shall denote by
the symbol $\Og{}$ the terms  proportional to the first degree
of
$g$
and by
$\Ogg$ --- the terms  proportional to the second and higher degrees of
$g$.

First of all, the respective approximation of the components of the FMT reads
as follows:
\be
g_{00}(g;R)=1-g\frac{\mR R^0}{(R^0)^2-\mR^2}-\frac12g\ln\frac{R^0-\mR}{R^0+\mR}+\Ogg,
\ee
\ses
\be
g_{0a}(g;R)=g\frac{\mR R^a}{(R^0)^2-\mR^2}+\Ogg,
\ee
\ses
\be
g_{ab}(g;R)=
-\delta_{ab}+
\frac12g\ln\frac{R^0-\mR}{R^0+\mR}\delta_{ab}-g\frac{R^aR^bR^0}{\mR((R^0)^2-\mR^2)}+
\Ogg;
\ee
\ses\\
\be
g^{00}(g;R)=1+g\frac{\mR R^0}{(R^0)^2-\mR^2}+\frac12g\ln\frac{R^0-\mR}{R^0+\mR}+\Ogg,
\ee
\ses
\be
g^{0a}(g;R)=g\frac{\mR R^a}{(R^0)^2-\mR^2}+\Ogg,
\ee
\ses
\be
g^{ab}(g;R)=
-\delta^{ab}-\frac12g\ln\frac{R^0-\mR}{R^0+\mR}\delta^{ab}+
g\frac{R^aR^bR^0}{\mR ((R^0)^2-\mR^2)}
+\Ogg.
\ee
The occurrence of the logarithmic function is characteristic of the approximations.
For the function (A.15) we get
\be
j(g;R)=
1-\frac14g\ln\frac{R^0-\mR}{R^0+\mR}
+\Ogg
\ee
so that, owing to (A.18),
\be
\det(g_{pq})=
-1+2g\ln\frac{R^0-\mR}{R^0+\mR}
+\Ogg.
\ee


The structure of the tensor
$
C_{pqr}
$ is such that the tensor is proportional to
$g$, so that
\be
C_{pqr}=gS_{pqr}+O_{\ge2}(g),
\ee
where the components $S_{pqr}$ are given by the list
\be
R^0S_{000}=\frac{R^0\mR^3}{((R^0)^2-\mR^2)^2},\qquad
R^0S_{a00}=-\frac{(R^0)^2\mR R^a}{((R^0)^2-\mR^2)^2},
\ee
\ses
\be
R^0S_{ab0}=
\frac12\frac{R^0\mR}{(R^0)^2-\mR^2}\delta_{ab}+
\frac12\frac{R^0R^aR^b((R^0)^2+\mR^2)}{\mR((R^0)^2-\mR^2)^2},
\ee
\ses
\be
R^0S_{abc}=
-\frac12\frac{(R^0)^2}{\mR((R^0)^2-\mR^2)}(\delta_{ab}R^c+\delta_{ac}R^b+\delta_{bc}R^a)+
\frac{(R^0)^2R^aR^bR^c((R^0)^2-3\mR^2)}{2\mR^3((R^0)^2-\mR^2)^2}.
\ee
For the components
\be
S_p{}^q{}_r=e^{qs}S_{psr},\qquad
S^{pqr}=e^{ps}e^{qt}e^{rv}S_{ptv},
\ee
we obviously obtain
\be
S_0{}^0{}_0=S_{000},\qquad
S_0{}^a{}_0=-S_{0a0},\qquad
S_a{}^0{}_0=-S_{a00},\qquad S_a{}^b{}_c=-S_{abc},
\ee
and
\be
S^{000}=S_{000},\qquad S^{a00}=-S_{a00},\qquad
S^{ab0}=S_{ab0},\qquad
S^{abc}=-S_{abc}.
\ee

For contractions $C_p=\3Cpqq$ we get
\be
C_p=gS_p+\Ogg,\quad C^p=gS^p+\Ogg,
\ee
and
\be
R^0S_0=-\frac N2\frac{R^0\mR}{(R^0)^2-\mR^2},\qquad R^0S_a=\frac
N2\frac{(R^0)^2R^a}{\mR((R^0)^2-\mR^2)},
\ee
\ses
\be
S^0=S_0,\qquad S^a=-S_a.
\ee
Obviously
\be
C_pC^p=\Ogg.
\ee
Also,
\be
\partial_pC^p=-\fr N2g(N-2)\frac{R^0}{\mR((R^0)^2-\mR^2)}
\ee
($(N=4)$).

The above corrections are singular over the light cone, that is, when
$R^0-\mR=0$.
The field $\{S_p\}$ introduces a preferred vector over the cone.
The tensor $\{S_{pqr}\}$
assigns an anisotropic structure over the background pseudoeuclidean space and enters
 respective covariant derivatives.


At the $O_1(g)$--level of consideration the curvature tensor of the space under study
is zero:
\be
S_{pqrs}=0
+\Ogg
\ee
(see (A.22) and (A.23)).
The same conclusion is applicable to the quasi-pseudoeuclidean objects:
\be
n_{ij}(g;t)=e_{ij}
+\Ogg,
\qquad
n^{ij}=e^{ij}
+\Ogg,
\ee
\ses
\be
\3Nikj(g;t)
=0
+\Ogg,
\qquad
R_{mnij}(g;t)=
0
+\Ogg
\ee
(see  (B.16), (B.26), and (B.34)).
Also,
the conformal properties are trivial at the $O_1(g)$--level of consideration:
\be
\ga=\Ogg
\ee
(see (B.37)).

\bigskip

\def\bibit[#1]#2\par{\rm\noindent\parskip1pt
                     \parbox[t]{.05\textwidth}{\mbox{}\hfill[#1]}\hfill
                     \parbox[t]{.925\textwidth}{\baselineskip11pt#2}\par}

\nin{\bf References}
\bigskip

\bibit[1] H.~ Busemann: \it Canad. J. Math. \bf1 \rm(1949), 279.

\bibit[2] H.~ Rund: \it The Differential Geometry of Finsler spaces, \rm
Springer-Verlag, Berlin 1959.



\bibit[3] G.S.~ Asanov: \it Rep. Math. Phys. \bf 45 \rm(2000), 155;
\bf 47 \rm(2001), 323.

\bibit[4] G.S.~ Asanov:arXiv:hep-ph/0306023, 2003;
arXiv:math.MG/0402013, 2004;
arXiv:math-ph/0406029, 2004.

\bibit[5] G.S.~ Asanov: \it Publ. Math. Debrecen \bf 61 \rm(2005), 1.

\bibit[6] G.S.~ Asanov:arXiv:gr-qc/0204070, 2002.

\end{document}